\documentclass{aa}  

\usepackage{graphicx}
\usepackage{txfonts}

\def\ms{\hbox{\,m\,s$^{-1}$}}         
\def\m2s2{\hbox{\,m$^{2}$\,s$^{-2}$}} 
\def\kms{\hbox{\,km\,s$^{-1}$}}       
\def\vsini{\hbox{$v$\,sin\,$i$}}      
\def\sini{\hbox{sin\,$i$}}      
\def\Msun{\hbox{$M_{\odot}$}}             

\def\Mjup{\hbox{$\mathrm{M}_{\rm Jup}$}}

\def \1s{$1\,\sigma$}

\def \t0{T$_0$}

\def \sophie{{\it SOPHIE}}
\def\modif{}

\usepackage{amstext}

\begin{document} 

\title{The $SOPHIE$ search for northern extrasolar planets\thanks{Based on observations made with 
ELODIE and $SOPHIE$ spectrographs on the 1.93-m telescope at Observatoire de Haute-Provence (CNRS/AMU), 
France. }}

   \subtitle{VIII. Follow-up of ELODIE candidates: long-period brown-dwarf companions\thanks{Tables 5 to 9 are available in electronic form
at the CDS via anonymous ftp to cdsarc.u-strasbg.fr (130.79.128.5)
or via http://cdsweb.u-strasbg.fr/cgi-bin/qcat?J/A+A/
} }

\author{
F. Bouchy \inst{1,2}
\and D. S\'egransan \inst{2}
\and R.F. D\'iaz \inst{1,2}
\and T. Forveille \inst{3} 
\and I. Boisse \inst{1,4}
\and L. Arnold \inst{5}
\and N. Astudillo-Defru \inst{3}
\and J.-L. Beuzit \inst{3}
\and X.~Bonfils \inst{3}
\and S. Borgniet \inst{3}
\and V. Bourrier \inst{2,6}
\and B. Courcol \inst{1}
\and X. Delfosse \inst{3}
\and O. Demangeon \inst{1}
\and P. Delorme \inst{3}
\and D. Ehrenreich \inst{2}
\and G.~H\'ebrard \inst{5,6}
\and A.-M. Lagrange \inst{3}
\and M. Mayor \inst{2}
\and G. Montagnier \inst{5,6}
\and C. Moutou \inst{1, 7} 
\and D. Naef \inst{2}
\and F. Pepe \inst{2}
\and C. Perrier \inst{3}
\and D.~Queloz \inst{2}
\and J. Rey \inst{2}
\and J. Sahlmann \inst{8}
\and A. Santerne \inst{4}
\and N.C. Santos \inst{4,9}
\and J.-P. Sivan \inst{1}
\and S. Udry \inst{2}
\and P.A. Wilson \inst{6}
}

\institute{
Aix Marseille Universit\'e, CNRS, Laboratoire d'Astrophysique de Marseille UMR 7326, 13388 Marseille cedex 13, France
\and
Observatoire de Gen\`eve, Universit\'e de Gen\`eve, 51 Ch. des Maillettes, 1290 Sauverny, Switzerland
\and  
Universit\'e Grenoble Alpes, CNRS, IPAG, 38000 Grenoble, France 
\and
Instituto de Astrof\'isica e Ci\^encias do Espa\c{c}o, Universidade do Porto, CAUP, 4150-762 Porto, Portugal
\and
Aix Marseille Universit\'e, CNRS, Observatoire de Haute-Provence, Institut Pyth\'eas, 04870 St Michel l'Observatoire, France
\and
Institut d'Astrophysique de Paris, UMR 7095 CNRS, Universit\'e Pierre \& Marie Curie, 98bis Bd Arago, 75014 Paris, France
\and
Canada France Hawaii Telescope Corporation, Kamuela 96743, USA 
\and
European Space Agency, European Space Astronomy Centre, P.O. Box 78, Villanueva de la Canada, 28691 Madrid, Spain
\and
Departamento de F\'isica e Astronomia, Faculdade de Ci\^encias, Universidade do Porto, 4150-762 Porto, Portugal 
}

   \date{Received ; accepted }

 
  \abstract
{Long-period brown dwarf companions detected in radial velocity surveys are important targets for direct imaging and astrometry to calibrate the mass -- luminosity relation of substellar objects. Through a 20-year radial velocity monitoring of solar-type stars that began with ELODIE and was extended with {\sophie} spectrographs, giant exoplanets and brown dwarfs with orbital periods longer than ten years are discovered. We report the detection of five new potential brown dwarfs with minimum masses between 32 and {\modif 83} {\Mjup} orbiting solar-type stars with periods longer than ten years. An upper mass limit of these companions is provided using astrometric Hipparcos data, 
high-angular resolution imaging made with PUEO, and a deep analysis of the cross-correlation function of the
main stellar spectra to search for blend effects or faint secondary components. These objects double the number of known brown dwarf companions with orbital periods longer than ten years and reinforce the conclusion that the occurrence of such objects increases with orbital separation. With a projected separation larger than 100 mas, all these brown dwarf candidates are appropriate targets for high-contrast and high angular resolution imaging.}


   \maketitle
%

\section{Introduction}

Brown dwarfs (BD) are substellar objects in the mass range of approximately 13 - 80 Jupiter masses;
they have enough mass to burn deuterium, but are too light to permit hydrogen 
burning in their inner cores \citep{1997ApJ...491..856B,2000ARA&A..38..337C,2011ApJ...727...57S}. 
BDs constitute intermediate objects between giant planets and low-mass stars, 
but there is not a complete consensus about their formation mechanisms. 
The separation between planet and brown dwarf population may not only be related to the mass, 
but also to the formation scenario. Core-accretion models predict the formation of objects as heavy 
as 25 {\Mjup} \citep{2009A&A...501.1161M}.   
BD companions orbiting solar-type stars with periods of up to $\text{about ten}$ years clearly occur
less frequently ($\le$1\%) than planetary systems and stellar binaries \citep[e.g.,][]{2006ApJ...640.1051G, 
2011A&A...525A..95S,2014MNRAS.439.2781M}. Statistical properties of BD companions, such as frequency, 
separation, eccentricity, and mass ratio distribution, as well as the relation of these properties to their host stars, 
should permit us to distinguish between different formation and evolution models. 

Most of the brown dwarf companions have been discovered in radial 
velocity (RV) surveys \citep[e.g.,][]{2002ApJS..141..503N,2011A&A...525A..95S,2012A&A...538A.113D}. 
In these programs, which were designed to find exoplanets, brown dwarfs are easily 
detected by their strong RV signatures. However, radial velocity measurements 
alone do not {\modif constrain} the orbit inclination and therefore provide only the minimum mass. 
When available, complementary observational constraints, such as the astrometric motion of the host star, 
a deep analysis of the spectra to search for a blend or a secondary component, or a photometric transit, 
are required to attempt determining the true mass or at least to firmly exclude the stellar nature of the companion.   

So far, only a few brown-dwarfs have been detected in radial velocity with orbital periods longer 
than ten years; this is a somewhat biased result because there
have been relatively few long-term surveys. These objects are important 
targets for direct imaging and astrometry with the aim to calibrate the mass -- luminosity relation of substellar objects.
Since the end of 2006, the {\sophie} consortium conducts a large program to search for exoplanets \citep{2009A&A...505..853B}. One of the different subprograms focuses on following-up the drifts and long-period signals detected within the ELODIE historical program initiated on the 1.93 m telescope at the Observatoire de Haute-Provence (OHP) by M. Mayor and D. Queloz in 1994. This program has monitored about 400 solar-type stars (from F5 to K0 dwarfs) in 12 years and led to the detection of 24 exoplanets \cite[e.g.,][]{1995Natur.378..355M,2003A&A...410.1039P,2004A&A...414..351N}. In the past years, this program was focused on monitoring RV drifts that were identified as incomplete orbits of gravitationally bound companions. Since 2007, we have continued this program with the {\sophie} spectrograph with the objective to characterize orbital parameters of exoplanets similar to Jupiter and Saturn in our solar system and brown dwarfs with orbital periods longer than eight to ten years.  
About 50 targets with a slow RV drift were selected from the original ELODIE sample and are followed-up with {\sophie}. This unique sample 
allows extending the time base to more than 20 years to find substellar companions beyond 5 AU.   
As part of this subprogram, we reported the detection of two Jupiter-analogs around HD150706 
and HD222155, we refined the parameters of the long-period planets HD154345b and HD89307b, 
and we determined the first reliable orbit for HD24040b \citep{2012A&A...545A..55B}.

In this paper, we report the detection of five new brown dwarf candidates with minimum masses of between 
32 and 78 {\Mjup} orbiting solar-type stars with period longer than ten years. We attempt to derive an upper limit on the true 
mass of these companions to determine their nature using 
astrometric Hipparcos data, high-angular resolution imaging made with PUEO for two of our targets, and deep analysis of the cross-correlation function (CCF) of our spectra to search for a blend effect or a faint secondary component. Such a CCF analysis was performed on \object{HD16702} and allowed identifying the secondary component.  
We also present in the Appendix an update of the orbital parameter of the low-mass star \object{HD29461}b that was recently published by \citet{2012JApA...33...29G} as well as the long-term RV drift of \object{HD175225}, which is induced by a low-mass stellar companion.


\section{Spectroscopic observations}

The targets reported in this paper and their basic stellar characteristics are listed in Table~\ref{targets}. 
The parallaxes are derived from the new Hipparcos 
reduction \citep{2007A&A...474..653V}. The stellar rotation parameter {\vsini} is derived from the calibration of the {\sophie} cross-correlation 
function by \citet{2010A&A...523A..88B} with an uncertainty of $\sim$ 1 {\kms}. The time span of the observations and the numbers of the ELODIE and {\sophie} measurements are listed in Table~\ref{targets}.

Spectroscopic observations of our targets were first conducted with the ELODIE spectrograph 
\citep{1996A&AS..119..373B} mounted on the 1.93 m telescope at OHP between late 1993 and mid-2006. The stars were then monitored by the {\sophie} 
spectrograph \citep{2008SPIE.7014E..0JP} from early 2007 to late 2014. 
The spectrograph configuration, the observing mode, and the data reduction were the same as 
those described in \citet{2012A&A...545A..55B}. 

One of the main {\sophie} systematic effects was caused by the insufficient scrambling of the fiber link and 
the high sensitivity of the spectrograph to illumination variation. Seeing change at the fiber input introduced 
radial velocity changes. This so-called seeing effect on {\sophie} has been described by \citet{2010A&A...523A..88B,2011A&A...528A...4B}, \citet{2012A&A...538A.113D}, and \citet{2013A&A...549A..49B}. Considering that variations in the pupil illumination produce a differential 
RV effect along spectral orders, the difference between RVs measured 
in each half of the spectral orders can be used to partially correct the main RV. We systematically applied the 
correction of the seeing effect by computing the RV difference between the left part (or blue side) and right part 
(or red side) of the spectral orders and using the average correlation coefficient of -1.04 estimated on standard stars 
\citep{2013A&A...549A..49B}. We added quadratically a systematic RV error of 6 {\ms} on the {\sophie} measurements 
corresponding to the average RV RMS measured on these stars. 
In June 2011 (BJD=2455730), a piece of octagonal-section fiber was implemented in the {\sophie} fiber link and significantly improved the radial velocity precision by eliminating the seeing effect, see \cite{2013A&A...549A..49B}.  
We checked on standard stars that the RV offset introduced by the instrumental upgrade
is smaller than the systematic errors of {\sophie}.     
Since this change and despite a clear improvement in RV precision on times scales of a few tens of days, 
some intervention within the instrument and strong variations of the outside temperature
propagated to the tripod of the spectrograph introduced zero-point offsets. The systematic drift 
was corrected using a set of RV constant star and following the approach described by {\modif \citet{2015A&A...581A..38C} }. 
  
\begin{table*}
\caption{Target characteristics and summary of observations.}           
\label{targets}      
\centering                          
\begin{tabular}{l c c c c c c c c c}        
\hline\hline                 
Target &  RA  &  DEC & V  & B-V   & Spectral  &  \vsini        &  Distance  & Time Span  &      Nmeas  \\
 Name  &        [deg]   & [deg]  &       &        &   Type      & [km/s]          &  [pc]    &    [years]        &  Elodie/Sophie \\
\hline                        
\object{HD10844}   & 26.687  &   25.918  &   8.13  &   0.63  &  F8V   &  1.9$\pm$1    & 52.3  &   12.78   &   25/27    \\      
\object{HD14348}   &  34.971  & 31.337   & 7.19   &    0.60    & F5V  &   5.4$\pm$1     & 56.6  &   16.85  &  61/38   \\      
\object{HD18757}  & 46.040  & 61.706  &  6.64   &  0.63      & G4V   &   1.4$\pm$1   & 24.2   &  16.75  & 31/43   \\   
\object{HD72946} &  128.964 & 6.623  &  7.25  &    0.71   &  G5V   &   3.9$\pm$1    & 26.2   &   16.04  &  45/23   \\ 
\object{HD209262} & 330.476  & 4.770  &   8.00  &   0.69     & G5V  &  3.0$\pm$1     & 49.7  &  12.70   & 15/21   \\  
\hline                                   
\end{tabular}
\end{table*}

\section{Data analysis}

\subsection{Stellar characteristics}

Spectra obtained without simultaneous thorium-argon calibration and with signal-to-noise ratios in the range of 150-200 were used to accurately derive the stellar effective temperature $T_{eff}$ , surface gravity log\,$g$, and metallicity [Fe/H]. The spectroscopic analysis method is described in \citet{2004A&A...415.1153S} and \citet{2008A&A...487..373S}. In brief, the procedure is based on the equivalent widths of the FeI and FeII lines and the iron excitation and ionization equilibrium, which is assumed to be in local thermodynamic equilibrium. Using the derived spectroscopic parameters as input, stellar masses $M_\star$ were derived from the calibration of \citet{2010A&ARv..18...67T} with a correction following \citet{2013A&A...556A.150S}. Errors were computed from 10 000 random values of the stellar parameters within their error bars and assuming a Gaussian distribution. 
Ages were estimated using the method\footnote{http://stev.oapd.inaf.it/param} described by \citet{2006A&A...458..609D} using PARSEC stellar tracks and isochrones from \citet{2012MNRAS.427..127B}.  
The stellar activity level was estimated from the emission in the core of the Ca II H and K bands measured on each SOPHIE spectra with the calibration described in \citet{2010A&A...523A..88B}. The average values derived from this analysis are reported in Table~\ref{spectro} 
with an estimated uncertainty of 0.05.

\begin{table*}
\caption{Stellar parameters}           
\label{spectro}      
\centering                          
\begin{tabular}{l c c c c c c c}        
\hline\hline                 
Target &  $T_{eff}$   & log\,$g$ & [Fe/H]  & $v$micro & $M_\star$  & Age   & $\log{R'_\mathrm{HK}}$ \\
 Name  &          [K]     &  [cgs]           &   [dex]    & [\kms]     &    [\Msun]  &   [Gyr]  & \\         
\hline                        
\object{HD10844}   & 5845$\pm$37 & 4.43$\pm$0.05 & -0.06$\pm$0.03 & 1.08$\pm$0.05   & 0.98$\pm$0.07    &  6.7$\pm$1.5  &   -4.85$\pm$0.05  \\      
\object{HD14348}   & 6237$\pm$47  & 4.51$\pm$0.07 & 0.28$\pm$0.03 & 1.53$\pm$0.06   &  1.20$\pm$0.08    & 2.2$\pm$0.3    & -4.95$\pm$0.05 \\   
\object{HD18757}   & 5656$\pm$28 & 4.41$\pm$0.04 & -0.27$\pm$0.02 & 0.84$\pm$0.04   &  0.88$\pm$0.06   &  11.4$\pm$0.1   &  -4.94$\pm$0.05 \\     
\object{HD72946}   & 5686$\pm$40 & 4.50$\pm$0.06 & 0.11$\pm$0.03  & 0.83$\pm$0.06   &  0.96$\pm$0.07    &  2.0$\pm$1.7   & -4.74$\pm$0.05 \\      
\object{HD209262} & 5815$\pm$28  & 4.41$\pm$0.04 & 0.10$\pm$0.02 & 1.14$\pm$0.03 & 1.02$\pm$0.07    & 4.7$\pm$1.8    &  -4.97$\pm$0.05  \\        
\hline                                   
\end{tabular}
\end{table*}

\subsection{Radial velocity analysis and orbital solutions}

ELODIE and {\sophie} radial velocities were fit using the Bayesian genetic software {\it Yorbit} {\modif , which is briefly described in \citet{2011A&A...535A..54S}. 
{\modif
The  main  advantage  of  a genetic  algorithm compared  to  other  advanced  methods  such  as a Markov  chain
Monte Carlo method (hereafter MCMC)  with parallel tempering is that it allows probing the full model parameter space in an extremely efficient way. 
The first step of the radial velocity data analysis consists of identifying significant periodic signals in the data. This is done using the General Lomb-Scargle periodogram algorithm \citep{2009A&A...496..577Z}, which is applied to the radial velocity measurements. False-alarm probabilities are estimated through a bootstrap approach by permuting the data. When\ a significant peak is located at a given period, the corresponding Keplerian is adjusted. 
For multiple Keplerians, the adjusted component is removed, the process is repeated several times until no significant peak remains, and all 
parameters are readjusted at each step of the analysis. 
However, the population at the end of the evolution is not statistically reliable owing the intrinsic nature of genetic algorithm, which
is based on genome 
crossover and mutation. To obtain robust confidence intervals for the Keplerian parameters as well as an estimate of the additional noise present in the data (called nuisance parameter), we further probed the parameter space with an MCMC algorithm with  a Metropolis-Hastings
algorithm. 
This is based on the formalism described in \cite{2007MNRAS.380.1230C} with  several  thousand chains drawn from the final genetic algorithm population. Each chain runs several thousand times to retrieve a statistically  reliable population. For poorly constrained planetary
systems, the Bayesian formalism described in \cite{2007MNRAS.374.1321G} is used. 
Our MCMC probes the following set of parameters: log $P$, $\sqrt{e}$\,cos$\omega$, $\sqrt{e}$\,sin$\omega$, log $K$, epoch of periastron $T_p$ or epoch of minimum RV $T_{RVmin}$, global RV offset ($V_0$ + $Ke$\,cos$\omega$) using {\sophie} as reference, and the  RV offset between ELODIE and {\sophie} ($\Delta V_{ES}$). The noise model follows a simple normal law with a standard deviation derived from the observation errors and a nuisance parameter ($\sigma^{2}_{tot} = \sigma^{2} + s^{2}$). Jeffrey priors are used for the period, the radial
velocity semi-amplitude, and the nuisance parameter, while uniform priors are used for the other parameters.
}

The RV offset between the two instruments was described and calibrated by \citet{2012A&A...545A..55B}. It is a function of the $B-V$ index and 
the cross-correlation mask used.  However, considering the quite large uncertainty of the RV offset relation (23 {\ms} RMS), we decided to let this parameter free to vary for the Keplerian fit.  For all the candidates we also tested a model including a long-term drift in addition to the Keplerian, 
but no significant drift was found, or it was strongly correlated to the ELODIE-{\sophie} offset. 
No significant additional signals were found on the RV residuals for the five targets. The orbital solutions are listed 
in Table~\ref{fit}, and each candidate is discussed in Sect.~\ref{bd}. \\

\subsection{Diagnostic for resolving the nature of the companions} 

The RV analysis provides a minimum mass for the companion within the range of brown dwarf masses. 
We used other diagnostics to attempt to derive an upper limit on the mass. 
More specifically, we used Hipparcos astrometry, high angular resolution imaging, and bisector span and CCF diagnostics.

\subsubsection{Hipparcos astrometry} 

For the five BD candidates,  we used the new Hipparcos reduction \citep{2007A&A...474..653V} to search 
for signatures of orbital motion in the Intermediate Astrometric Data (IAD). The analysis was performed 
as described in \citet{2011A&A...525A..95S} using the orbital elements given by the RV analysis and reported in Table~\ref{fit}. 
The basic parameters of the Hipparcos observations relevant for the astrometric analysis 
are listed in Table~\ref{hipp}. 

\begin{table}
\caption{Parameters of the Hipparcos astrometric observations}           
\label{hipp}      
\centering                          
\begin{tabular}{l c c c c}        
\hline\hline                 
Target &   HIP  &   Nmeas    &   Time Span  &   $\sigma$  \\
 Name  &                  &                   &   [days]  &    [mas]    \\         
\hline                        
\object{HD10844}   & 8285   & 94   &   1069  &   3.34     \\      
\object{HD14348}     & 10868   & 82   &  1096   &  1.69         \\       
\object{HD18757}    & 14286  & 129  &  1043  &   2.59         \\ 
\object{HD72946}    & 42173  &  61  &    759  &  1.71      \\    
\object{HD209262} &   108761  &  57  &  1084  &  3.23     \\     
\hline                                   
\end{tabular}
\end{table}

For the five BD candidates we found no detectable signal of a corresponding astrometric motion in the Hipparcos data. The main reason is that the orbital periods are very long and the astrometry data cover only $3-23$\% of the orbital phase. This inhibits a significant detection even though the expected minimum signals are quite strong (4-24 mas) compared to the measurement precision. Furthermore, it cannot be excluded that most of the strong general signal is absorbed into a slightly incorrect proper motion.

\subsubsection{Bisector span analysis} 

Following the approach described by \citet{2012A&A...538A.113D}, we examined the bisector span of the {\sophie} CCF of our BD candidates to check whether significant variation and a correlation 
with RV measurements can reveal a relatively luminous companion polluting the peak 
of the primary star. The detection of a correlation indicates that the velocity signal is the result - at least in part - 
of a deformation of the stellar CCF. For the five targets the bisector 
span varies only negligibly and does not exhibit any significant correlation with radial velocity 
measurements; this excludes massive M dwarfs. 
The correlation slopes were systematically smaller than 8 {\ms} per {\kms} within the uncertainties.  
However, such a non-detection does not definitively exclude low-mass stellar companions. 
Simulations indicate that only massive M dwarfs heavier than 0.5 $M_\sun$ can be firmly excluded. \\

 \subsubsection{Search for a second component in the CCF} 
 
If the companion is relatively bright, we expect that it is blended within the main spectrum and then pollutes the CCF of the primary star.   
To search for a secondary peak in our {\sophie} CCFs, we proceeded in the follow way: first we built the CCFs 
using an M mask to maximize the contribution of a suspected low-mass star companion. Then we shifted all the CCFs 
to the systemic velocity $V_0$ and averaged them, after normalization of the minimum and maximum of the CCF, to build 
a master CCF. We then subtracted this master CCF from all the individual CCFs shifted to the systemic velocity. The velocity of the second 
component $Vc$ is expected to be at $ - V_1/q$ with respect to the systemic velocity $V_0$, with $V_1$ the velocity of the main component and $q$ the mass ratio $Mc/M_\star$. Considering that all individual CCFs are shifted to the systemic velocity, this second component is expected 
to be at $(V_0 - V_1) \times (1 + 1/q)$. To increase the detection capability, we combined all the CCF residuals after shifting them   
to this last quantity with $q$ varying from 0.1 to 0.6. For each value of $q$ we computed the dispersion of the average of CCF residuals 
and searched for a significant peak centered at the systemic velocity. 

We tested our approach on \object{HD16702}, which was first identified by \citet{2012A&A...538A.113D} to have a companion with a minimum mass of 48.7 {\Mjup}. These authors showed that Hipparcos astrometric analysis and CCF simulations lead to a companion mass of 0.55 $\pm$ 0.14 $M_\odot$ and 0.40 $\pm$ 0.05 $M_\odot$ , respectively. In the average CCF residuals, shown in Fig.~\ref{hd16702}, we find a deepest secondary peak of 1.3 $\pm$ 0.1 \% when a shift of $ (3.5 \pm 0.4) \times (V_0 - V_1) $ is imposed, hence for a mass ratio $q=0.4 \pm 0.06$. The uncertainty on the shift applied to have the deepest second peak was estimated considering the noise level within the continuum (0.1\%) and determining the shift limit, which produced a peak depth of 
1.2\%. 

Given a primary mass of 0.98 $M_\odot$, the companion is then an M2-type star of 0.39 $\pm$ 0.06 $M_\odot$ in agreement with \citet{2012A&A...538A.113D}. This value also agrees with the identification of a second component in the spectra 
by \citet{2015AJ....149...18K} with an estimated $T_{eff}$ of 3500 $\pm$ 250 K and contributing 
to 1.06\% of the total flux. More recently, {\modif \citet{2015MNRAS.451.2337S}} performed blend models with PASTIS  \citep{2014MNRAS.441..983D} and derived a companion mass of 0.35 $\pm$ 0.03 $M_{\odot}$.  

We estimated that for \object{HD16702} the detection limit is close to 0.3\%.  
However, this approach requires several observations well spread across the orbit, and the systemic velocity 
should be known to correctly stack and average the CCFs. Furthermore, the effectiveness 
of the approach strongly depends on the orbital parameters, the
number of measurements, and the phase coverage. 

We applied this search for a second component on our candidates, but without any significant detection. 
To estimate the effectiveness of the approach and derive an upper limit, we injected an artificial second 
component into the CCFs that corresponds to a companion with a velocity at $Vc+(Vc-V_1)/q$, a CCF contrast from 0.2 to 2\%  of the primary, 
and the same CCF FWHM as the primary. We note that a larger CCF FWHM of the secondary, although not expected for low-mass stars of several billion years, reduces the detection sensitivity.
The detection limit obtained a range from 0.5\% up to 2\% for the worst cases like \object{HD10844} and 
\object{HD209262}, for which the {\sophie} velocities only cover one side with respect to the systemic velocity. In this case the second component is not 
well averaged and not enough diluted when computing the master CCF. It was therefore partially removed in all the individual CCFs when we subtracted the master CCF. Assuming that the CCF contrast ratio could be used as a rough estimate of the flux ratio between primary and secondary, 
we derive an upper mass limit using the evolutionary models for the main sequence of low-mass stars by {\modif  \citet{2015A&A...577A..42B}} taking into account the mass and age of the primary. The derived upper limits are in the range of 0.18-0.4 $M_{\odot}$ and are reported in Table~\ref{fit}. 
The lowest upper limit of 0.18 $M_{\odot}$ is found for \object{HD18757}.

\begin{figure}
\centering
\includegraphics[width=\hsize]{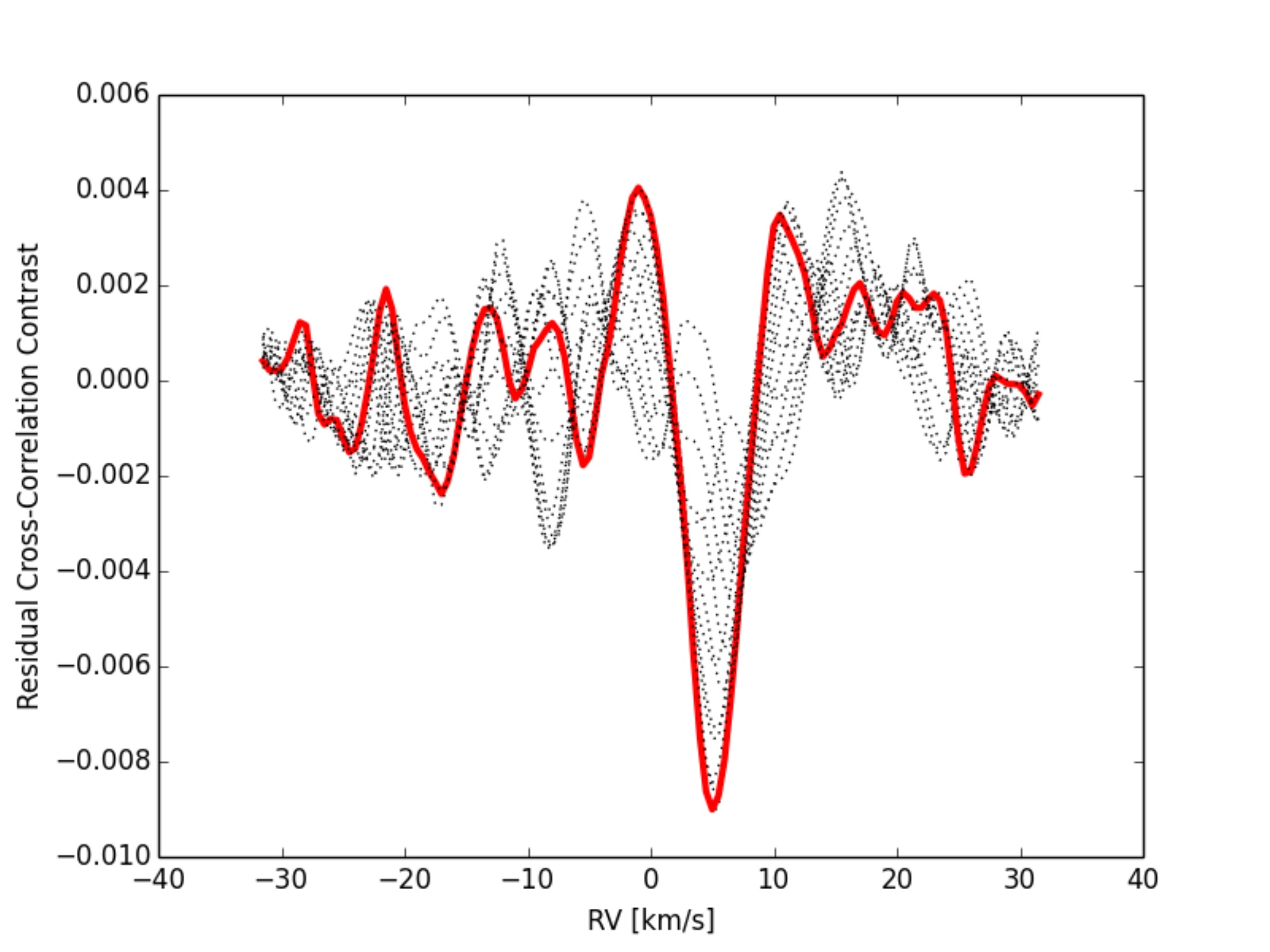}
\caption{Averaged CCF residuals of \object{HD16702} for different radial velocity shifts. The deepest peak (red curve) occurs for a shift corresponding to a mass ratio of 0.4.}
\label{hd16702}
\end{figure}

\subsubsection{High-resolution imaging with PUEO}  

\object{HD14348} and \object{HD18757} were observed as filler targets during a program dedicated to mass measurements for M dwarfs 
at the 3.6 m Canada-France-Hawaii Telescope (CFHT) using PUEO, the CFHT Adaptive Optics Bonnette \citep{1998PASP..110..152R},
and the KIR infrared camera \citep{1998SPIE.3354..760D}. The observing procedure is described by \citet{1999A&A...344..897D}. 
High angular resolution imaging for these two stars was analyzed to derive a detection limit for the companion. 
The flux detection limit was converted into an upper mass limit considering the age of the primary and using the models NEXTGEN for low-mass stars 
\citep{1998A&A...337..403B} and DUSTY for brown dwarfs \citep{2001ApJ...556..357A}.   
The RV orbital parameters and the mass of the primary were used to compute the sky-projected orbital separation of the 
companion relative to the host star at the date of PUEO observations assuming different orbital inclinations. 

Figure~\ref{figpueo} shows the mass detection limit as function of the angular separation for \object{HD18757}. The green line represents the 
projected separation varying the orbital inclination. Companions along the green line and above black line (NEXTGEN) or red line (DUSTY) 
are excluded. This gives an upper limit of 0.13 $M_\odot$ for \object{HD18757}b, which is lower than the upper limit derived from 
the CCF analysis. 

For \object{HD14348}, the upper limit is 0.6 $M_\odot$, which is well above the upper limit derived from the CCF analysis because 
the angular separation is close to 0.1 arcsec, just below the diffraction limit of the CFHT in K band (0.12 arcsec). \\

\begin{figure}
\centering
\includegraphics[width=8cm]{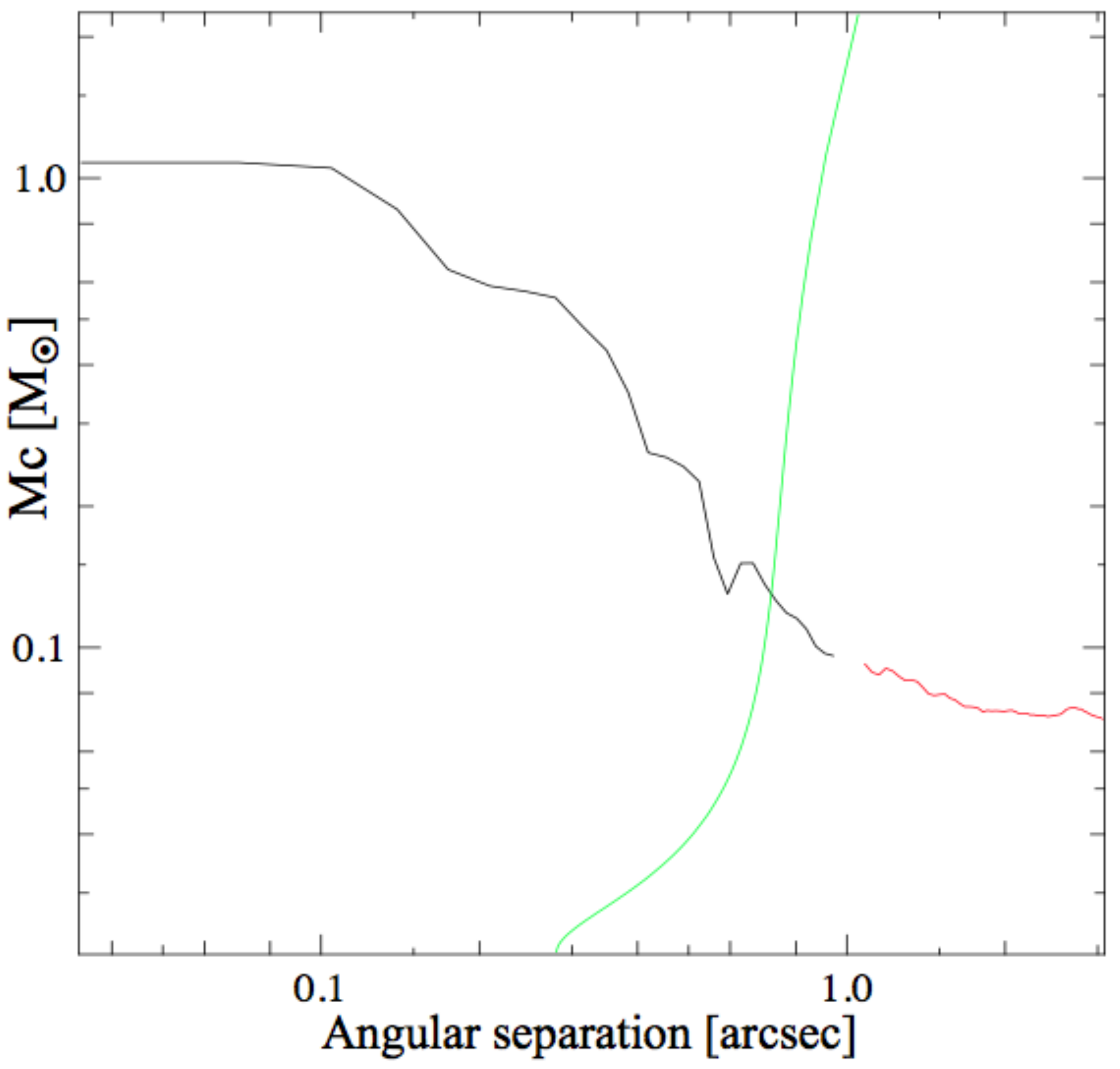}
\caption{Mass detection limit as function of angular separation for \object{HD18757} observed with PUEO. The black and red curves correspond to NEXTGEN and DUSTY models, respectively. The green line represents the 
projected separation varying the orbital inclination and shows that no companion with a mass higher than 0.13 $M_{\odot}$ was detected. }
\label{figpueo}
\end{figure}

\section{Long-period brown dwarf companions}
\label{bd}

\begin{table*}
\caption{{\modif Median and 1-$\sigma$ limit of posterior parameters derived from our MCMC analysis.}}           
\label{fit}      
\centering                          
\begin{tabular}{l c c c c c c}        
\hline\hline                 
Parameters  &   \object{HD10844}       &  \object{HD14348}     &  \object{HD18757}                      &  \object{HD72946}                & \object{HD209262}  \\
\hline               
P [years]   & 31.6$^{+2.7}_{-2.1}$ & 12.987$^{+0.017}_{-0.013}$     & 109$^{+18}_{-16}$        & 15.93$^{+0.15}_{-0.13}$      & 14.88$^{+0.37}_{-0.26}$   \\[+3pt]
K [\ms]     &  873$^{+54}_{-35}$       & 575.0$^{+2.2}_{-2.6}$       &  665.8$^{+5.9}_{-5.3}$             & 776$\pm$9.0       &  385$^{+13}_{-11}$ \\[+3pt]
e             & 0.568$^{+0.022}_{-0.020}$ & 0.455$\pm$0.004   & 0.943$\pm$0.007       & 0.495$\pm$0.006  &  0.347$^{+0.010}_{-0.008}$ \\[+3pt]
$\omega$ [deg]  & -94.9$\pm$1.4   & 65.12$^{+0.55}_{-0.64}$       & 44.0$^{+0.5}_{-0.4}$               & 71.0$\pm$1.6          &  -104.1$^{+3.1}_{-3.5}$ \\[+3pt]
$T_p$ -2450000      & 3876.7$\pm$16   & 5265.6$\pm$4.5   &  5220.0$\pm$0.8     &   5958$\pm$10         &  6766$\pm$6 \\[+3pt]
$V_0$ [\kms] & -42.708$^{+0.035}_{-0.054}$  & -4.231$^{+2.2}_{-2.6}$  & -2.201$^{+0.006}_{-0.005}$  & 29.527$\pm$0.008  &  -43.329$^{+0.013}_{-0.012}$ \\[+3pt]
$\Delta V_{ES}$ [\kms] & -98$^{+104}_{-70}$ & -57$\pm$3.7     &  -120$^{+9.5}_{-8.0}$         &   -113$\pm$14      &  -165$\pm$24  \\[+3pt]
Jitter $s$ (E/S) [\ms]   &  17.8/1.2      & 14.8/5.3            & 1.4/2.7                           &  21.8/15.5             &  14.3/3.3 \\[+3pt]
$a$ [AU]                  & 10.18$^{+0.62}_{-0.49}$   & 5.95$\pm$0.10    &   22.2$^{+2.5}_{-2.3}$          & 6.37$\pm$0.11    & 6.16$^{+0.15}_{-0.13}$   \\[+3pt]
$a$ [mas]                & 195$^{+12}_{-9}$      &   105.1$\pm$1.7     & 917$^{+103}_{-95}$           &  243.1$\pm$4.2              &   123.9$^{+3.0}_{-2.6}$    \\[+3pt]
Mc\,\sini [\Mjup]     &   83.4$^{+6.4}_{-5.0}$   &   48.9$\pm$1.6  &    35.2$\pm$1.2   & 60.4$\pm$2.2    & 32.3$^{+1.6}_{-1.5}$  \\[+3pt]
Mc,upper-lim [$M_{\odot}$] & 0.4           &  0.35             &   0.13               &       0.2                       &   0.4      \\
\hline                                   
\end{tabular}
\end{table*}

\subsection{\object{HD10844}}

\object{HD10844} is a V = 8.13 mag F8V star with a mass $M_\star$ of 0.98 $M_\odot$ located at 52.3 parsec  from the Sun. The Keplerian fit 
to the RV measurements, listed in Table~\ref{rv10844}, indicates a companion of minimum mass $Mc\,${\sini} = {\modif 83.4} {\Mjup} on a 30-year orbit, with an eccentricity $e$ = 0.57. With such a minimum mass, \object{HD10844}b is fairly likely to be stellar, since the inclination only needs to be slightly less than edge-on.
Although the measurement time-span covers less than half of the orbital period of \object{HD10844}, the Keplerian fit 
is quite well constrained because the orbit is eccentric and has a large semi-amplitude.  
The instrumental offset between ELODIE and {\sophie} is adjusted to {\modif -98 $^{+104}_{-70}$}  {\ms} in agreement with the expected one at -66 $\pm$ 23 {\ms}. {\modif Using the expected instrumental offset as prior does not significantly change the result, but it reduces the uncertainties with P = 32.3$^{+1.0}_{-1.2}$ years and K = 885$^{+12}_{-11}$ {\ms}.} 
Systematic errors of 17.8 {\ms} and 1.2 {\ms} were quadratically added to ELODIE and {\sophie,} respectively, to obtain the reduced $\chi^2$ = 1. 

 \begin{figure}[h]
\centering
\includegraphics[width=8cm]{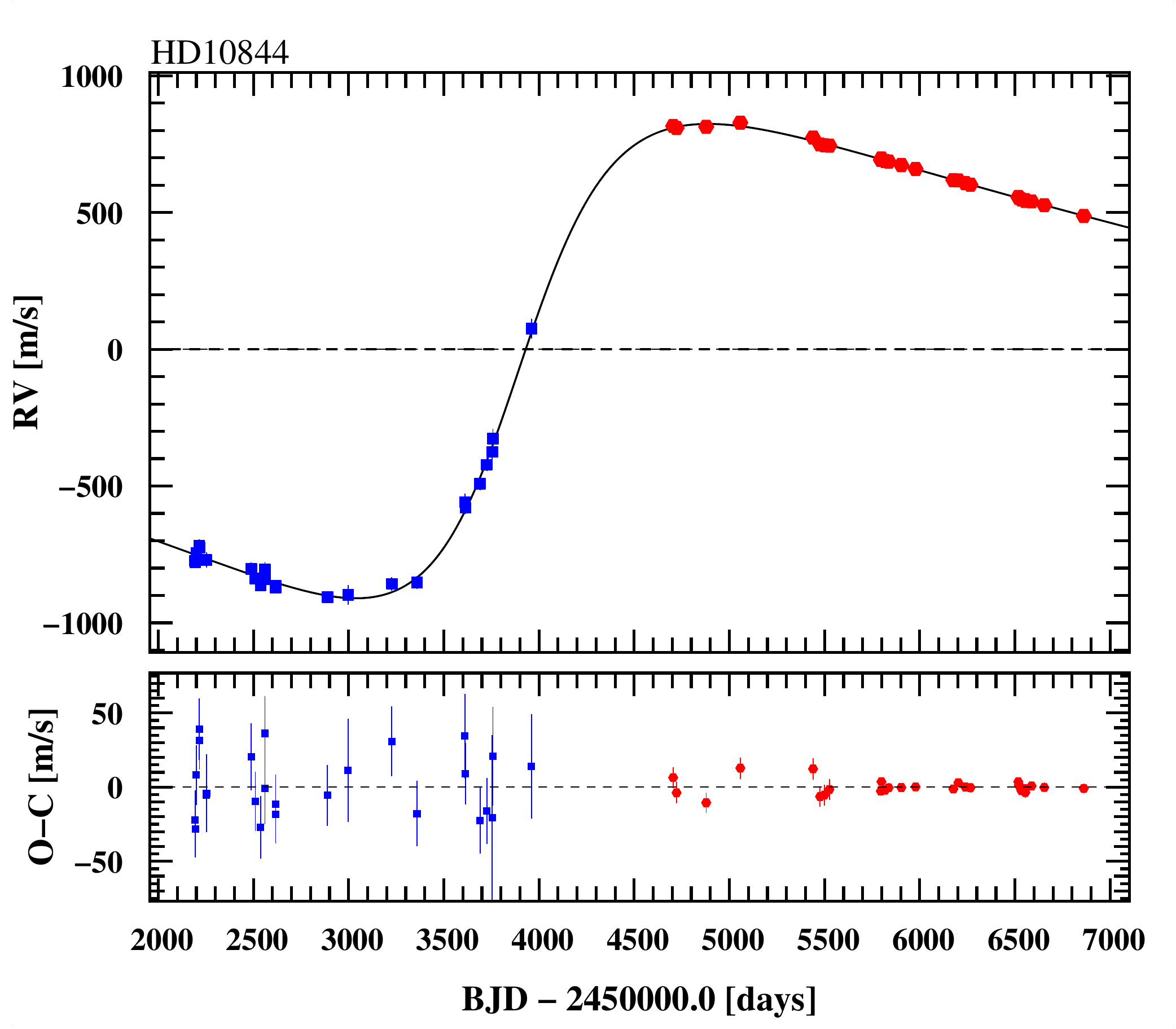}
\caption{Radial velocity curve of \object{HD10844} obtained with ELODIE (blue) and SOPHIE (red).}
\label{fighd10844}
\end{figure}

\subsection{\object{HD14348}} 

\object{HD14348} is a F5V star with V = 7.2 mag located at 56.6 parsec from the Sun, with a mass of 1.2 $M_\odot$. The RV measurements listed in Table~\ref{rv14348} show a peak-to-peak variation of 1150 {\ms} that reveal the presence of an $Mc\,${\sini} = {\modif 48.9} {Mjup} companion on a 13-year orbit, with an eccentricity of 0.45. 
The instrumental offset between ELODIE and {\sophie} is adjusted to -57 $\pm$ 3.7 {\ms} , which agrees 
with the expected offset at -53 $\pm$ 23 {\ms}. Systematic errors of 14.8 {\ms} and 5.3 {\ms} were quadratically added to ELODIE and {\sophie,} respectively. 

\begin{figure}[h]
\centering
\includegraphics[width=8cm]{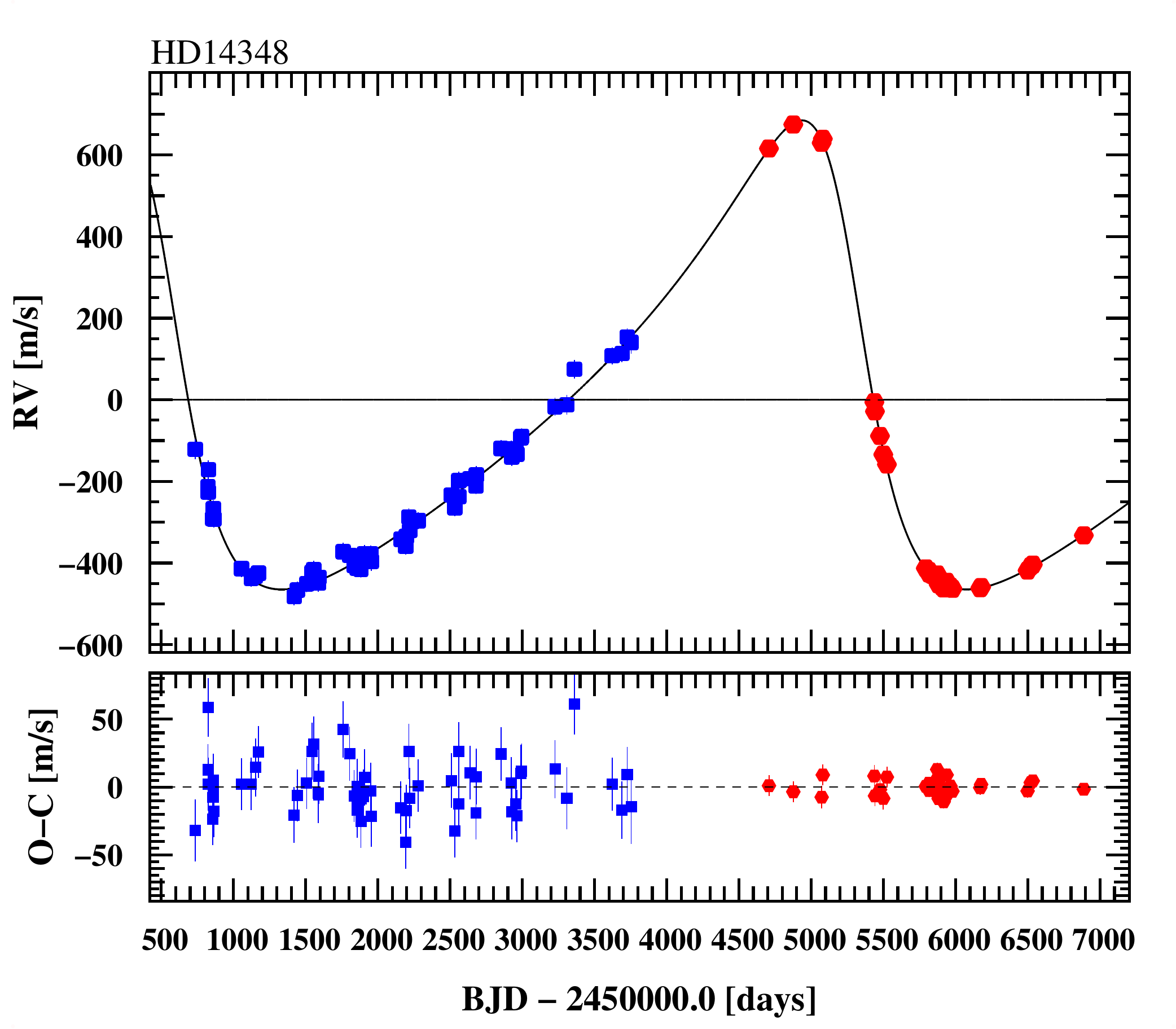}
\caption{Radial velocity curve of \object{HD14348}.}
\label{fighd14348}
\end{figure}

\subsection{\object{HD18757}}

\object{HD18757} is a bright V = 6.64 mag G4V star with low metallicity located at only 24.2 parsec from the Sun. 
\object{HD18757} is part of a multiple system with an M-dwarf component of V = 12.5 located at 12.7 arcsec (307 AU) \citep{2002yCat.1274....0D} 
and a physically associated common proper motion companion with a separation of 262 arcsec (6340 AU) and a spectral type of M2V \citep{2010ApJS..190....1R}. The RV measurements listed in Table~\ref{rv18757} are fitted by a 109-year Keplerian orbit with a semi-amplitude $K$ = 666 {\ms} and 
a high eccentricity $e$ = 0.94. The spectral analysis together
with the comparison with evolutionary models yields 
a stellar mass of $M_\star$ = 0.88 $M_\odot$ and a companion minimum mass of {\modif 35.2} {\Mjup}. 
Although a small fraction (17\%) of the orbital period is covered, the Keplerian fit is sufficiently well {\modif constrained}  because the orbit is highly eccentric and the RV extrema are covered.  
The instrumental offset between ELODIE and {\sophie} is adjusted to {\modif -120 $^{+9.5}_{-8.0}$} {\ms} , which is marginally  
higher than the expected offset at $-66 \pm 23$ {\ms}. Systematic errors of 1.4 {\ms} and 2.7 {\ms} were quadratically added to ELODIE and {\sophie,} respectively. 
The large distance of the companion and the vicinity of the target allows placing a quite strong constraint on 
the mass upper limit derived from PUEO observations with $Mc\,\le\,0.13$ $M_\odot$.  

\begin{figure}[h]
\centering
\includegraphics[width=8cm]{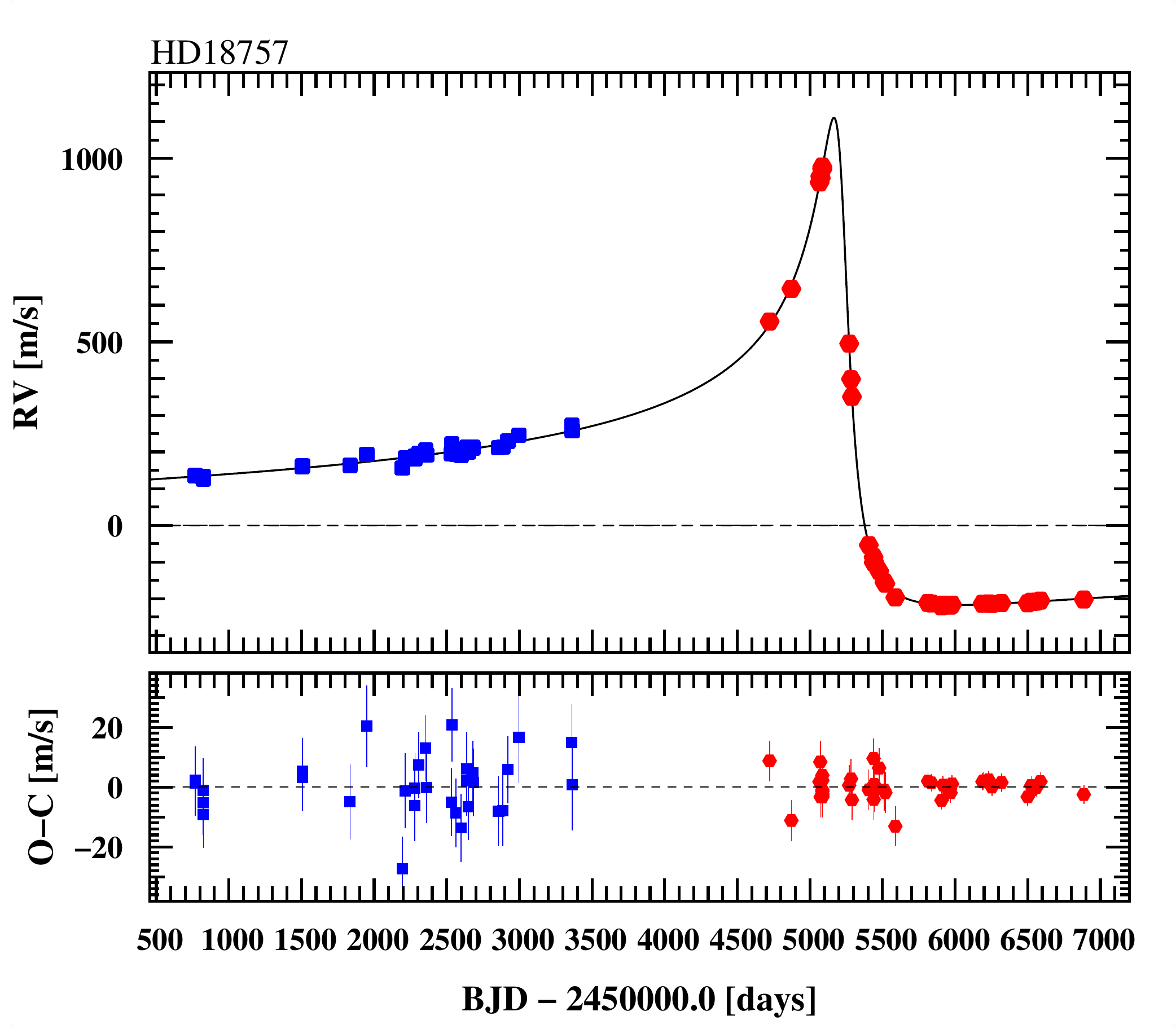}
\caption{Radial velocity curve of \object{HD18757}.}
\label{fighd18757}
\end{figure}

\subsection{\object{HD72946}}

\object{HD72946} is a V = 7.25 mag G5 star with a mass of 0.96 $M_\odot$ located at 26.2 parsec from the Sun. 
\object{HD72946} (GJ310.1B) has a physically associated common proper motion companion HD72945 (GJ310.1A) 
located at 10 arcsec (230 AU). This companion is a spectroscopic binary with a period of 14.3 days 
including an F8V primary of mv=5.99 \citep{1991A&A...248..485D}. 
We checked that this bright component located at 10 arcsec cannot introduce contaminating light 
within the 3-arcsec fiber acceptance of {\sophie}. Even with a seeing as high as 6 arcsec, the contamination remains  
lower than 0.3\% and therefore is negligible. 
\object{HD72946} is also part of a multiple stellar system with a V = 10.7 star at 93 arcsec 
and a V = 12 star at 117 arcsec \citep{2002yCat.1274....0D}. 
The Keplerian fit to the RV measurements listed in Table~\ref{rv72946} indicates a companion of minimum mass $Mc\,${\sini} = {\modif 60.4} {\Mjup} in a 16-year orbit, 
with an eccentricity $e$ = 0.495. The instrumental offset between ELODIE and {\sophie} is adjusted to -113 $\pm$ 14 {\ms} , which fully agrees with the expected offset of -100 $\pm$ 23 {\ms}.
Systematic errors of 22 {\ms} and 15 {\ms} were quadratically added to ELODIE and {\sophie,} respectively, 
which may be due to the fact that the star is slightly active ($\log{R'_\mathrm{HK}}$ = -4.74). 
Although a non-significant drift was found from fixing the instrument offset, we note that 
the F8V companion at 230 AU may introduce a radial velocity change on \object{HD72946} of up to few {\ms} per year.     

\begin{figure}[h]
\centering
\includegraphics[width=8cm]{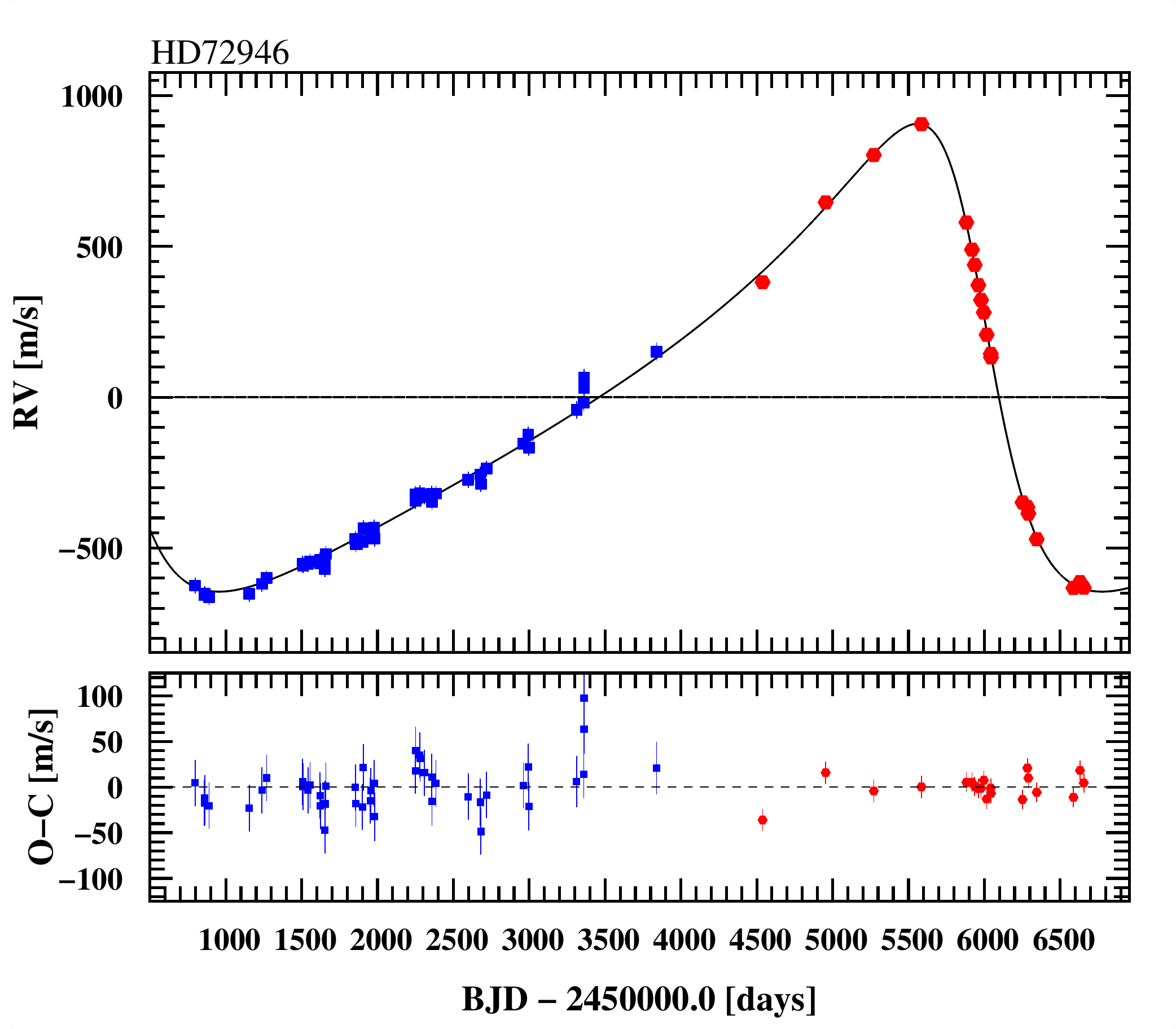}
\caption{Radial velocity curve of \object{HD72946}.}
\label{fighd72946}
\end{figure}

\subsection{\object{HD209262}}

\object{HD209262} is a V = 8.0 G-type star located at 49.7 parsec from the Sun. 
\object{HD209262} is part of a double system \citep{2002yCat.1274....0D} with a component BD+044788B (V = 9.8) at 78.3 arcsec  
and possibly a third component (V = 12.4) at 22.4 arcsec. 
The RV measurements listed in Table~\ref{rv209262} are fitted by a 15-year Keplerian orbit with a semi-amplitude of 385 {\ms} and 
an eccentricity $e$ = 0.35. The spectral analysis combined with the comparison with evolutionary models yields 
a stellar mass of $M_\star$ = 1.02 $M_\odot$ and a companion minimum mass of {\modif 32.3} {\Mjup}. 
The instrumental offset between ELODIE and {\sophie} is adjusted to {\modif -165 $\pm$ 24} {\ms} , which is marginally higher than 
the expected offset at -91 $\pm$ 23 {\ms}. Systematic errors of 14.3 {\ms} and 3.3 {\ms} were quadratically added 
to ELODIE and {\sophie,} respectively. 

\begin{figure}[h]
\centering
\includegraphics[width=8cm]{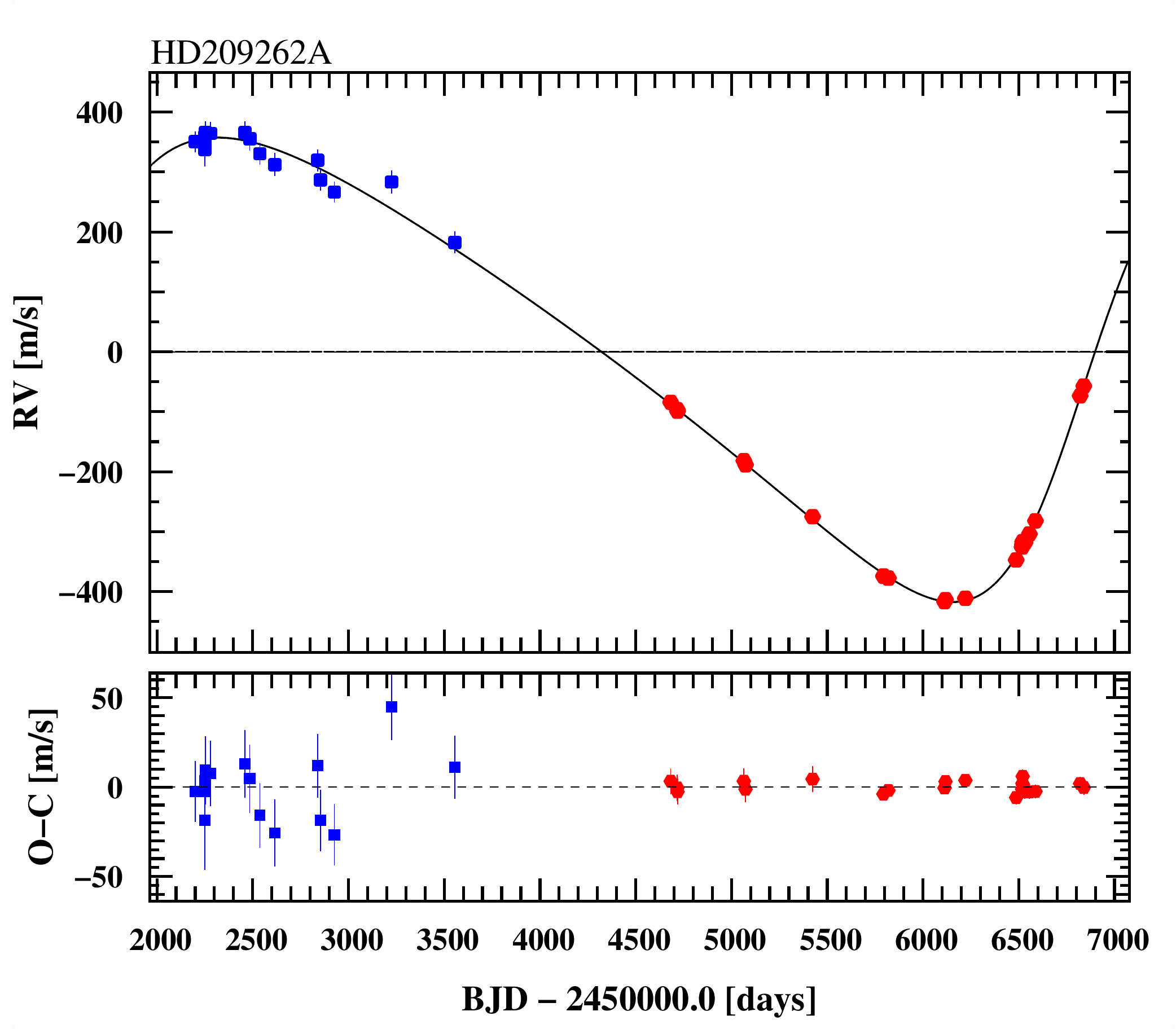}
\caption{Radial velocity curve of \object{HD209262}.}
\label{fighd208262}
\end{figure}

\section{Discussion and conclusion}

We have reported the detection of five new potential brown dwarf companions with minimum masses of between 32 and {\modif 83} {\Mjup} orbiting solar-type stars with periods longer than ten years. These detection were made thanks to ELODIE and {\sophie} 
radial velocity measurements that cover a time span of up to 16 years. Our different diagnostics allowed excluding companions 
heavier than 0.4 $M_{\odot}$ and for \object{HD18757} heavier than 0.13 $M_{\odot}$. 
These candidates double the number of known BD companions with orbital periods 
longer than ten years, which include \object{HD4747}, \object{HD74014}, \object{HD211847}, and \object{HD167665} \citep{2011A&A...525A..95S} that were detected in the CORALIE RV survey, 
and \object{HIP5158} \citep{2010A&A...512A..48L,2011MNRAS.416L.104F} from the HARPS RV survey. This last one, \object{HIP5158}b, should probably 
be considered a planetary candidate given its minimum mass of 15 $\pm$ 10 {\Mjup}.   
This increasing number of long-period BD companions reinforces the observation that the number of BDs increases with orbital 
period, as pointed out by \citet{2014MNRAS.439.2781M}. 
All our BD candidates have an eccentricity higher than 0.45 except
for the lightest one, \object{HD209262}b, with $e$ = 0.35. 
Three of our five targets are part of a multiple stellar system. 
All our targets, with a projected separation larger than 100 mas, and especially \object{HD18757} with {\modif 917 mas,} 
will be good targets for high-contrast and high angular resolution imaging, interferometry, or speckle observations.  
Such targets are important to constrain the mass -- luminosity relation of brown dwarfs for different ages and metallicity.  Micro-arcsecond astrometry with the Gaia satellite \citep{2011AdSpR..47..356M} will permit deriving constraints on the orbital inclination and the true mass of these companions.  

\begin{longtab}
\begin{longtable}{cccc}
\caption{\label{rv10844} Radial velocities of \object{HD10844}}\\
\hline\hline
BJD &  RV  &  $\sigma_{RV}$  & bisector span \\
-2\,450\,000  &  [{\kms}]   &   [{\kms}]   &  [{\kms}]  \\
\hline
\endfirsthead
\caption{continued.}\\
\hline\hline
BJD &  RV  &  $\sigma_{RV}$  & bisector span \\
-2\,450\,000  &  [{\kms}]   &   [{\kms}]   &  [{\kms}]  \\
\hline
\endhead
\hline
\endfoot
\hline                        
\multicolumn{4}{c}{ELODIE} \\
\hline
2193.5946 &     -43.5490 &      0.0100 & - \\
2194.5399 &     -43.5510 &      0.0090 & - \\
2199.5132 &     -43.5240 &      0.0110 & - \\
2215.4148 &     -43.5360 &      0.0120 & - \\
2216.4245 &     -43.5450 &      0.0100 & - \\
2253.4252 &     -43.5660 &      0.0150 & - \\ 
2253.4463 &     -43.5750 &      0.0200 & - \\
2488.6193 &     -43.5870 &      0.0150 & - \\
2509.6445 &     -43.6090 &      0.0110  & -\\
2537.6158 &     -43.6380 &      0.0130 & - \\
2560.4763 &     -43.6340 &      0.0170 & - \\
2560.4895 &     -43.6180 &      0.0190 & - \\
2616.4728 &     -43.6540 &      0.0100 & - \\
2616.4856 &     -43.6570 &      0.0100 & - \\
2888.6468 &     -43.6880 &      0.0120 & - \\
2997.3236 &     -43.6970 &      0.0310 & - \\
3226.6045 &     -43.6490 &      0.0170 & - \\
3359.3799 &     -43.6410 &      0.0150  & -\\
3610.6239 &     -43.3370 &      0.0230 & - \\
3613.6502 &     -43.3750 &      0.0120 & - \\
3690.4311 &     -43.3090 &      0.0140 & - \\
3726.2632 &     -43.2290 &      0.0150 & - \\
3754.3081 &     -43.1570 &      0.0550 & - \\
3758.2915 &     -43.1170 &      0.0290 & - \\
3960.6272 &     -42.7270 &      0.0320 & - \\
\hline                        
\multicolumn{4}{c}{SOPHIE} \\
\hline         
4704.6436 &     -41.8846 &      0.0064 &        -0.0182 \\
4722.5421 &     -41.8921 &      0.0065 &        -0.0065 \\
4878.2689 &     -41.8878 &      0.0063 &        -0.0103 \\
5057.5982 &     -41.8727 &      0.0066 &        -0.0250 \\
5439.6580 &     -41.9270 &      0.0065 &        -0.0228 \\
5476.5150 &     -41.9522 &      0.0061 &        -0.0105 \\
5499.3238 &     -41.9553 &      0.0063 &        -0.0152 \\
5525.4397 &     -41.9563 &      0.0064 &        -0.0050 \\
5795.6368 &     -42.0089 &      0.0015 &        -0.0137 \\
5798.6131 &     -42.0032 &      0.0018 &        -0.0117 \\
5817.4916 &     -42.0126 &      0.0017 &        -0.0150 \\ 
5836.4271 &     -42.0147 &      0.0018 &        -0.0147 \\
5903.2592 &     -42.0277 &      0.0020 &        -0.0088 \\
5978.3057 &     -42.0421 &      0.0018 &        -0.0130 \\
6176.5696 &     -42.0827 &      0.0016 &        -0.0178 \\
6202.5796 &     -42.0836 &      0.0016 &        -0.0120 \\
6239.4788 &     -42.0938 &      0.0017 &        -0.0132 \\ 
6266.2769 &     -42.0995 &      0.0022 &        -0.0145 \\
6517.6399 &     -42.1444 &      0.0018 &        -0.0135 \\
6520.6155 &     -42.1473 &      0.0018 &        -0.0133 \\
6521.6385 &     -42.1475 &      0.0018 &        -0.0117 \\
6533.5961 &     -42.1534 &      0.0020 &        -0.0173 \\
6554.5681 &     -42.1587 &      0.0024 &        -0.0187 \\
6555.5704 &     -42.1562 &      0.0017 &        -0.0138 \\
6587.4011 &     -42.1606 &      0.0019 &        -0.0193 \\
6654.3186 &     -42.1743 &      0.0021 &        -0.0138 \\
6861.6110 &     -42.2139 &      0.0023 &        -0.0128 \\    
\hline
\end{longtable}
\end{longtab}

\begin{longtab}
\begin{longtable}{cccc}
\caption{\label{rv14348} Radial velocities of \object{HD14348}}\\
\hline\hline
BJD &  RV  &  $\sigma_{RV}$  & bisector span \\
-2\,450\,000  &  [{\kms}]   &   [{\kms}]   &  [{\kms}]  \\
\hline
\endfirsthead
\caption{continued.}\\
\hline\hline
BJD &  RV  &  $\sigma_{RV}$  & bisector span \\
-2\,450\,000  &  [{\kms}]   &   [{\kms}]   &  [{\kms}]  \\
\hline
\endhead
\hline
\endfoot
\hline                        
\multicolumn{4}{c}{ELODIE} \\
\hline
50734.5038      &  -4.3840      &  0.0150  & - \\
50823.4128      &  -4.5170      &  0.0080  & - \\
50825.3663      &  -4.5250      &  0.0090  & - \\
50826.4153      &  -4.4710      &  0.0140  & - \\
50857.2826      &  -4.5740      &  0.0080  & - \\
50858.2950      &  -4.5620      &  0.0080  & - \\
50859.2953      &  -4.5610      &  0.0090  & - \\
50860.2912      &  -4.5590      &  0.0070  & - \\
50861.3098      &  -4.5640      &  0.0100  & - \\
50863.2966      &  -4.5770      &  0.0080  & - \\
51056.6372      &  -4.6750      &  0.0080  & - \\
51123.6089      &  -4.7190      &  0.0090  & - \\
51152.3989      &  -4.7010      &  0.0130  & - \\
51173.3453      &  -4.7130      &  0.0080  & - \\
51420.6399      &  -4.7360      &  0.0110  & - \\
51442.6286      &  -4.7270      &  0.0080  & - \\
51506.4760      &  -4.7340      &  0.0080  & - \\
51543.3230      &  -4.7110      &  0.0120  & - \\
51555.3205      &  -4.7090      &  0.0110  & - \\
51587.2820      &  -4.7250      &  0.0140  & - \\
51589.2960      &  -4.7250      &  0.0100  & - \\
51590.2757      &  -4.7250      &  0.0110  & - \\
51759.6206      &  -4.6810      &  0.0120  & - \\
51804.6178      &  -4.6550      &  0.0100  & - \\
51835.5704      &  -4.6730      &  0.0080  & - \\
51856.5246      &  -4.6940      &  0.0110  & - \\
51857.5569      &  -4.6880      &  0.0120  & - \\
51858.4078      &  -4.6760      &  0.0110  & - \\
51882.4306      &  -4.6870      &  0.0090  & - \\
51883.4170      &  -4.6730      &  0.0100  & - \\
51900.3237      &  -4.6770      &  0.0110  & - \\
51908.3544      &  -4.6520      &  0.0120  & - \\
51950.2663      &  -4.6680      &  0.0120  & - \\
51953.3094      &  -4.6840      &  0.0150  & - \\
52158.6712      &  -4.6130      &  0.0090  & - \\
52193.6244      &  -4.6020      &  0.0100  & - \\
52197.5729      &  -4.5770      &  0.0090  & - \\
52214.5084      &  -4.5580      &  0.0110  & - \\
52220.4888      &  -4.6000      &  0.0140  & - \\
52280.3416      &  -4.5800      &  0.0090  & - \\
52510.6264      &  -4.5270      &  0.0110  & - \\
52533.6126      &  -4.5140      &  0.0090  & - \\
52560.5083      &  -4.5030      &  0.0150  & - \\
52560.5214      &  -4.4510      &  0.0130  & - \\
52637.3375      &  -4.4910      &  0.0100  & - \\
52679.2899      &  -4.5070      &  0.0090  & - \\
52681.3245      &  -4.4780      &  0.0120  & - \\
52852.6229      &  -4.4200      &  0.0100  & - \\
52922.5259      &  -4.3840      &  0.0090  & - \\
52926.5355      &  -4.4040      &  0.0110  & - \\
52954.5217      &  -4.4040      &  0.0110  & - \\
52961.4165      &  -4.4150      &  0.0090  & - \\
52990.3551      &  -4.3860      &  0.0120  & - \\
52995.3754      &  -4.3760      &  0.0110  & - \\
53226.6196      &  -4.3330      &  0.0130  & - \\
53308.4870      &  -4.2830      &  0.0100  & - \\
53361.4090      &  -4.2110      &  0.0150  & - \\
53623.6143      &  -4.1670      &  0.0100  & - \\
53690.4781      &  -4.1730      &  0.0130  & - \\
53726.3125      &  -4.1330      &  0.0110  & - \\
53754.3416      &  -4.1220      &  0.0220  & - \\
\hline                        
\multicolumn{4}{c}{SOPHIE} \\
\hline         
54707.6566      & -3.6155       & 0.0061        & 0.0232  \\
54876.2743      & -3.5568       & 0.0063        & 0.0142  \\
55072.6061      & -3.6018       & 0.0068        & 0.0117  \\
55080.6299      & -3.5919       & 0.0062        & 0.0142  \\
55437.6260      & -4.2373       & 0.0064        & 0.0180  \\
55441.6044      & -4.2597       & 0.0062        & 0.0180  \\
55476.5474      & -4.3199       & 0.0061        & 0.0175  \\
55499.3410      & -4.3655       & 0.0066        & 0.0277  \\
55524.4661      & -4.3898       & 0.0063        & 0.0175  \\
55793.6268      & -4.6443       & 0.0017        & 0.0212  \\
55815.5634      & -4.6511       & 0.0015        & 0.0203  \\
55816.6297      & -4.6569       & 0.0019        & 0.0137  \\
55818.5053      & -4.6555       & 0.0015        & 0.0123  \\
55836.4801      & -4.6603       & 0.0015        & 0.0192  \\
55865.5665      & -4.6676       & 0.0020        & 0.0145  \\
55866.4560      & -4.6678       & 0.0017        & 0.0257  \\
55872.4819      & -4.6594       & 0.0024        & 0.0200  \\
55875.5262      & -4.6682       & 0.0019        & 0.0035  \\
55877.4609      & -4.6667       & 0.0019        & 0.0178  \\
55878.4741      & -4.6718       & 0.0019        & 0.0177  \\
55879.4537      & -4.6703       & 0.0018        & 0.0172  \\
55881.4784      & -4.6795       & 0.0019        & 0.0180  \\
55882.4706      & -4.6835       & 0.0017        & 0.0115  \\
55883.4788      & -4.6822       & 0.0018        & 0.0180  \\
55906.3028      & -4.6893       & 0.0018        & 0.0165  \\
55917.4365      & -4.6934       & 0.0025        & 0.0212  \\
55919.3907      & -4.6858       & 0.0020        & 0.0315  \\
55930.2903      & -4.6922       & 0.0018        & 0.0208  \\
55938.3275      & -4.6773       & 0.0017        & 0.0227  \\
55957.2527      & -4.6901       & 0.0029        & 0.0257  \\
55964.2788      & -4.6891       & 0.0021        & 0.0283  \\
55978.3194      & -4.6944       & 0.0017        & 0.0220  \\
56170.5917      & -4.6931       & 0.0029        & 0.0170  \\
56176.5812      & -4.6903       & 0.0015        & 0.0163  \\
56498.6138      & -4.6497       & 0.0019        & 0.0192  \\
56520.6342      & -4.6393       & 0.0016        & 0.0195  \\
56534.5942      & -4.6354       & 0.0015        & 0.0278  \\
56887.5803      & -4.5636       & 0.0014        & 0.0237  \\   
\hline
\end{longtable}
\end{longtab}

\begin{longtab}
\begin{longtable}{cccc}
\caption{\label{rv18757} Radial velocities of \object{HD18757}}\\
\hline\hline
BJD &  RV  &  $\sigma_{RV}$  & bisector span \\
-2\,450\,000  &  [{\kms}]   &   [{\kms}]   &  [{\kms}]  \\
\hline
\endfirsthead
\caption{continued.}\\
\hline\hline
BJD &  RV  &  $\sigma_{RV}$  & bisector span \\
-2\,450\,000  &  [{\kms}]   &   [{\kms}]   &  [{\kms}]  \\
\hline
\endhead
\hline
\endfoot
\hline                        
\multicolumn{4}{c}{ELODIE} \\
\hline
50767.4725      & -2.1700       & 0.0080 & - \\
50769.5273      & -2.1710       & 0.0080 & - \\
50822.3861      & -2.1800       & 0.0080 & - \\
50823.4239      & -2.1830       & 0.0070 & - \\
50824.3832      & -2.1850       & 0.0080 & - \\
51505.5103      & -2.1440       & 0.0080 & - \\
51506.4973      & -2.1430       & 0.0080 & - \\
51834.5794      & -2.1290       & 0.0100 & - \\
51950.3226      & -2.1230       & 0.0110 & - \\
52194.5828      & -2.1100       & 0.0070 & - \\
52214.5388      & -2.1100       & 0.0100 & - \\
52279.4027      & -2.1120       & 0.0090 & - \\
52281.3512      & -2.1080       & 0.0090 & - \\
52308.3416      & -2.1140       & 0.0080 & - \\
52355.2857      & -2.1060       & 0.0080 & - \\
52360.3018      & -2.1090       & 0.0090 & - \\
52531.6535      & -2.0760       & 0.0080 & - \\
52534.6050      & -2.0530       & 0.0100 & - \\
52561.6542      & -2.0940       & 0.0090 & - \\
52597.4496      & -2.1070       & 0.0090 & - \\
52637.3512      & -2.0920       & 0.0090 & - \\
52638.3956      & -2.0820       & 0.0100 & - \\
52649.3504      & -2.1040       & 0.0080 & - \\
52678.3316      & -2.0950       & 0.0070 & - \\
52682.3392      & -2.0970       & 0.0080 & - \\
52856.6065      & -2.0620       & 0.0090 & - \\
52885.6576      & -2.0760       & 0.0090 & - \\
52919.5345      & -2.0610       & 0.0080 & - \\
52996.3605      & -2.0610       & 0.0130 & - \\
53361.3514      & -2.0180       & 0.0110 & - \\
53363.4057      & -2.0390       & 0.0140 & - \\
\hline                        
\multicolumn{4}{c}{SOPHIE} \\
\hline   
54722.6450      & -1.6432       & 0.0061        & -0.0205 \\
54872.3146      & -1.5542       & 0.0062        & -0.0190 \\
55066.6401      & -1.2641       & 0.0062        & -0.0227 \\
55071.6531      & -1.2469       & 0.0062        & -0.0223 \\
55074.6245      & -1.2523       & 0.0061        & -0.0222 \\
55085.6225      & -1.2229       & 0.0061        & -0.0210 \\
55085.6316      & -1.2279       & 0.0061        & -0.0175 \\
55085.6394      & -1.2265       & 0.0061        & -0.0195 \\
55085.6467      & -1.2260       & 0.0061        & -0.0225 \\
55085.6539      & -1.2284       & 0.0061        & -0.0248 \\
55085.6617      & -1.2273       & 0.0061        & -0.0215 \\
55085.6690      & -1.2211       & 0.0061        & -0.0192 \\
55270.2770      & -1.7035       & 0.0061        & -0.0202 \\
55283.3034      & -1.8002       & 0.0061        & -0.0210 \\
55289.2925      & -1.8485       & 0.0061        & -0.0212 \\
55405.6056      & -2.2526       & 0.0061        & -0.0178 \\
55438.5861      & -2.2854       & 0.0061        & -0.0170 \\
55439.5806      & -2.3002       & 0.0061        & -0.0177 \\
55441.6413      & -2.2998       & 0.0062        & -0.0212 \\
55441.6469      & -2.2973       & 0.0062        & -0.0183 \\
55476.6276      & -2.3231       & 0.0061        & -0.0213 \\
55513.4199      & -2.3538       & 0.0061        & -0.0232 \\
55519.5212      & -2.3578       & 0.0061        & -0.0198 \\
55588.3009      & -2.3952       & 0.0061        & -0.0202 \\
55813.6358      & -2.4101       & 0.0012        & -0.0218 \\
55836.5278      & -2.4117       & 0.0011        & -0.0207 \\
55906.3411      & -2.4195       & 0.0014        & -0.0290 \\ 
55917.4699      & -2.4146       & 0.0017        & -0.0262 \\ 
55957.3354      & -2.4171       & 0.0017        & -0.0175 \\ 
55964.3435      & -2.4165       & 0.0015        & -0.0218 \\ 
55968.2763      & -2.4177       & 0.0019        & -0.0232 \\ 
55978.2971      & -2.4148       & 0.0013        & -0.0215 \\ 
56188.5902      & -2.4125       & 0.0012        & -0.0232 \\ 
56193.6382      & -2.4121       & 0.0013        & -0.0225 \\ 
56230.5639      & -2.4112       & 0.0013        & -0.0215 \\ 
56252.5244      & -2.4132       & 0.0012        & -0.0242 \\ 
56318.3870      & -2.4104       & 0.0015        & -0.0268 \\ 
56499.6173      & -2.4111       & 0.0015        & -0.0210 \\ 
56523.6283      & -2.4067       & 0.0013        & -0.0197 \\ 
56534.6057      & -2.4079       & 0.0012        & -0.0250 \\ 
56558.5126      & -2.4066       & 0.0013        & -0.0235 \\ 
56586.6204      & -2.4039       & 0.0014        & -0.0202 \\ 
56885.5818      & -2.4009       & 0.0015        & -0.0268 \\  
\hline
\end{longtable}
\end{longtab}

\begin{longtab}
\begin{longtable}{cccc}
\caption{\label{rv72946} Radial velocities of \object{HD72946}}\\
\hline\hline
BJD &  RV  &  $\sigma_{RV}$  & bisector span \\
-2\,450\,000  &  [{\kms}]   &   [{\kms}]   &  [{\kms}]  \\
\hline
\endfirsthead
\caption{continued.}\\
\hline\hline
BJD &  RV  &  $\sigma_{RV}$  & bisector span \\
-2\,450\,000  &  [{\kms}]   &   [{\kms}]   &  [{\kms}]  \\
\hline
\endhead
\hline
\endfoot
\hline                        
\multicolumn{4}{c}{ELODIE} \\
\hline
50797.6592      & 28.8010       & 0.0090 & - \\
50858.4937      & 28.7830       & 0.0080 & - \\
50860.4688      & 28.7820       & 0.0070 & - \\
50861.4710      & 28.7910       & 0.0080 & - \\
50889.4530      & 28.7610       & 0.0090 & - \\
51154.6532      & 28.7790       & 0.0100 & - \\
51238.4794      & 28.8210       & 0.0090 & - \\
51269.3387      & 28.8230       & 0.0080 & - \\
51506.6238      & 28.8660       & 0.0090 & - \\
51508.5795      & 28.8600       & 0.0090 & - \\
51541.5551      & 28.8690       & 0.0090 & - \\
51555.5717      & 28.8720       & 0.0080 & - \\
51621.3412      & 28.8790       & 0.0090 & - \\
51623.3434      & 28.8690       & 0.0090 & - \\
51653.3511      & 28.8590       & 0.0090 & - \\
51654.3364      & 28.8840       & 0.0080 & - \\
51659.3353      & 28.9050       & 0.0090 & - \\
51853.6795      & 28.9500       & 0.0090 & - \\
51856.6575      & 28.9380       & 0.0090 & - \\
51900.5962      & 28.9670       & 0.0080 & - \\
51906.6132      & 29.0010       & 0.0100 & - \\
51952.4446      & 28.9800       & 0.0080 & - \\
51956.4750      & 28.9950       & 0.0080 & - \\
51978.3723      & 29.0060       & 0.0100 & - \\
51980.4137      & 28.9700       & 0.0120 & - \\
52250.6469      & 29.0770       & 0.0090 & - \\
52250.6584      & 29.1000       & 0.0120 & - \\
52278.5817      & 29.1000       & 0.0080 & - \\
52281.5065      & 29.1020       & 0.0100 & - \\
52308.5537      & 29.1030       & 0.0090 & - \\
52358.3662      & 29.1060       & 0.0100 & - \\
52359.3811      & 29.0790       & 0.0130 & - \\
52385.3412      & 29.1130       & 0.0100 & - \\
52597.6778      & 29.1490       & 0.0090 & - \\
52677.4908      & 29.1770       & 0.0100 & - \\
52681.5314      & 29.1450       & 0.0080 & - \\
52719.3760      & 29.1810       & 0.0090 & - \\
52961.6485      & 29.2650       & 0.0080 & - \\
52993.6046      & 29.3040       & 0.0100 & - \\
52997.5657      & 29.2540       & 0.0110 & - \\
53312.6837      & 29.3860       & 0.0110 & - \\
53359.6308      & 29.4110       & 0.0110 & - \\
53361.6161      & 29.4590       & 0.0130 & - \\
53361.6286      & 29.5070       & 0.0130 & - \\
53839.3236      & 29.6250       & 0.0170 & - \\
\hline                        
\multicolumn{4}{c}{SOPHIE} \\
\hline    
54538.3770      & 29.9066       & 0.0062        & -0.0058 \\
54954.3455      & 30.1713       & 0.0062        & -0.0167  \\
55271.4482      & 30.3282       & 0.0061        & -0.0198 \\
55585.4876      & 30.4308       & 0.0061        & -0.0123 \\
55881.6709      & 30.1049       & 0.0020        & 0.0087 \\
55918.6887      & 30.0147       & 0.0021        & -0.0123 \\
55936.6008      & 29.9638       & 0.0017        & -0.0115 \\
55960.5143      & 29.8970       & 0.0020        & -0.0023 \\
55978.5115      & 29.8477       & 0.0021        & -0.0128 \\
55996.4528      & 29.8062       & 0.0015        & -0.0148 \\
56015.3346      & 29.7325       & 0.0016        & -0.0073 \\
56042.3515      & 29.6689       & 0.0015        & -0.0203 \\
56044.3510      & 29.6575       & 0.0016        & -0.0058 \\
56251.6506      & 29.1764       & 0.0013        & -0.0123 \\
56283.6289      & 29.1609       & 0.0019        & -0.0048 \\
56290.5846      & 29.1398       & 0.0015        & -0.0085 \\
56345.4729      & 29.0544       & 0.0017        & -0.0032 \\
56587.6983      & 28.8930       & 0.0023        & 0.0022 \\
56630.6525      & 28.9122       & 0.0013        & -0.0152 \\
56655.6067      & 28.8940       & 0.0016        & 0.0123 \\     
\hline
\end{longtable}
\end{longtab}

\begin{longtab}
\begin{longtable}{cccc}
\caption{\label{rv209262} Radial velocities of \object{HD209262}}\\
\hline\hline
BJD &  RV  &  $\sigma_{RV}$  & bisector span \\
-2\,450\,000  &  [{\kms}]   &   [{\kms}]   &  [{\kms}]  \\
\hline
\endfirsthead
\caption{continued.}\\
\hline\hline
BJD &  RV  &  $\sigma_{RV}$  & bisector span \\
-2\,450\,000  &  [{\kms}]   &   [{\kms}]   &  [{\kms}]  \\
\hline
\endhead
\hline
\endfoot
\hline                        
\multicolumn{4}{c}{ELODIE} \\
\hline
52199.3623      & -43.1070      & 0.0090 & - \\
52250.2869      & -43.1340      & 0.0240 & - \\
52251.2263      & -43.1300      & 0.0090 & - \\
52251.2380      & -43.1350      & 0.0110 & - \\ 
52252.2565      & -43.1140      & 0.0130 & - \\
52280.2394      & -43.1190      & 0.0120 & - \\
52459.5922      & -43.0990      & 0.0120 & - \\
52484.5762      & -43.1380      & 0.0130 & - \\
52537.4647      & -43.1720      & 0.0120 & - \\
52616.2435      & -43.1650      & 0.0120 & - \\
52838.5756      & -43.1640      & 0.0110 & - \\
52853.5755      & -43.2010      & 0.0090 & - \\
52926.4085      & -43.2050      & 0.0100 & - \\
53224.5707      & -43.2410      & 0.0120 & - \\
53555.6024      & -43.3080      & 0.0100 & - \\
\hline                        
\multicolumn{4}{c}{SOPHIE} \\
\hline         
54682.5576      & -43.4135      & 0.0062        & -0.0242 \\
54717.4614      & -43.4259      & 0.0065        & -0.0298 \\
54718.4912      & -43.4283      & 0.0063        & -0.0357 \\
55064.4498      & -43.5109      & 0.0062        & -0.0205 \\
55073.5169      & -43.5178      & 0.0064        & -0.0195 \\
55423.5417      & -43.6043      & 0.0063        & -0.0217 \\
55793.5068      & -43.7034      & 0.0016        & -0.0160 \\
55819.3782      & -43.7068      & 0.0013        & -0.0218 \\
56112.5968      & -43.7461      & 0.0015        & -0.0262 \\
56117.5850      & -43.7428      & 0.0016        & -0.0223 \\
56221.2753      & -43.7405      & 0.0015        & -0.0210 \\
56486.5924      & -43.6765      & 0.0019        & -0.0187 \\
56517.4967      & -43.6549      & 0.0017        & -0.0240 \\
56519.5470      & -43.6505      & 0.0019        & -0.0187 \\
56520.5602      & -43.6460      & 0.0016        & -0.0195 \\
56521.5195      & -43.6502      & 0.0018        & -0.0233 \\
56532.3651      & -43.6476      & 0.0022        & -0.0262 \\
56555.4319      & -43.6332      & 0.0021        & -0.0247 \\
56587.3708      & -43.6112      & 0.0017        & -0.0203 \\
56820.5924      & -43.4027      & 0.0020        & -0.0195 \\  
56839.5684      & -43.3864      & 0.0022        & -0.0285 \\   
\hline
\end{longtable}
\end{longtab}


\begin{acknowledgements}
We gratefully acknowledge the Programme National de Plan{\'e}tologie (telescope time attribution and financial support) of CNRS/INSU, the Swiss National Science Foundation, and the Agence Nationale de la Recherche (grant ANR-08-JCJC-0102-01) for their support. We warmly thank the OHP staff for their support on the 1.93 m telescope. J.S. is supported by an ESA Research Fellowship in Space Science. 
A.S. is supported by the European Union under a Marie Curie Intra-European Fellowship for Career Development with reference FP7-PEOPLE-2013-IEF, number 627202.
{\modif This work has been carried out in the frame of the National Centre for Competence in Research ``Planet'' supported by the Swiss National Science Foundation (SNSF). D.S., R.F.D., N.A., V.B., D.E., F.P., D.Q. and S.U. acknowledge the financial support of the SNSF. P.A.W acknowledges the support of the French Agence Nationale de la Recherche (ANR), under program ANR-12-BS05-0012 "Exo-Atmos". The Porto group acknowledges the support from Funda\c{c}\~ao para a Ci\^encia e a Tecnologia (FCT, Portugal) in the form of grants reference SFRH/BPD/70574/2010 and PTDC/FIS-AST/1526/2014. NCS also acknowledges the support from FCT in the form of grant reference PTDC/CTE-AST/098528/2008 and through Investigador FCT contract of reference IF/00169/2012 as well as POPH/FSE (EC) by FEDER funding through the program ``Programa Operacional de Factores de Competitividade - COMPETE''.} 
\end{acknowledgements}

\bibliographystyle{aa}
\bibliography{26347_ref}

\appendix 

\section{\object{HD29461}}

\object{HD29461} is a V = 7.96 mag G5-type star located at 46.5 parsec from our Sun. 
It was first identified as a spectroscopic binary in the Hyades field by \citet{1988AJ.....96..172G} 
with a period of about ten years. \citet{1993AJ....105..220M} and \citet{1998AJ....115.1972P} failed to resolve 
the system by speckle interferometry, indicating that the companion is very faint 
with a mass upper limit of 0.22 $M_\odot$. \citet{2007ApJ...665..744P} gave a preliminary orbit for \object{HD29461} 
from Keck observations that did not cover a complete cycle. The period they selected was 18 years. 
More recently, \citet{2012JApA...33...29G} published the orbital elements derived from 78 radial velocity 
measurements spanning 38 years, from 1972 to 2009. The Keplerian fit yields a 
period P = 3760$\pm$8 days, a semi-amplitude K = 1.36$\pm$0.04 {\kms} , and an eccentricity $e$ = 0.594$\pm$0.019. 

For this target we obtained 20 ELODIE and 20 {\sophie} measurements in a time span of 13 years. 
Our measurements, displayed in Fig.~\ref{fighd29461}, confirm and agree with the orbital 
parameters derived by \citet{2012JApA...33...29G},
except for our semi-amplitude (K = 1.483 $\pm$ 0.015 {\kms}), which is 3 $\sigma$ higher.   
The instrumental offset between ELODIE and {\sophie} is fitted to -167 $\pm$ 16 {\ms} , which is marginally 
higher than the expected offset at -78 $\pm$ 23 {\ms}. This slightly higher offset can explain the 
larger semi-amplitude found. Systematic errors of 15 {\ms} and 8 {\ms} were quadratically added to ELODIE and {\sophie,} respectively. 
The star is slightly active with $\log{R'_\mathrm{HK}}$=-4.73, which may explain the RV jitter observed 
in {\sophie} measurements. We derive a period P = 3776$\pm$12 days and an eccentric $e$ = 0.613$\pm$0.004. 
The spectral analysis and the comparison with evolutionary models yields 
a stellar mass of $M_\star$ = 1.06$\pm$0.07 $M_\odot$ and a companion minimum mass of 93.4 {\Mjup} 
, which firmly excludes a brown-dwarf companion. 

\begin{figure}[h]
\centering
\includegraphics[width=8cm]{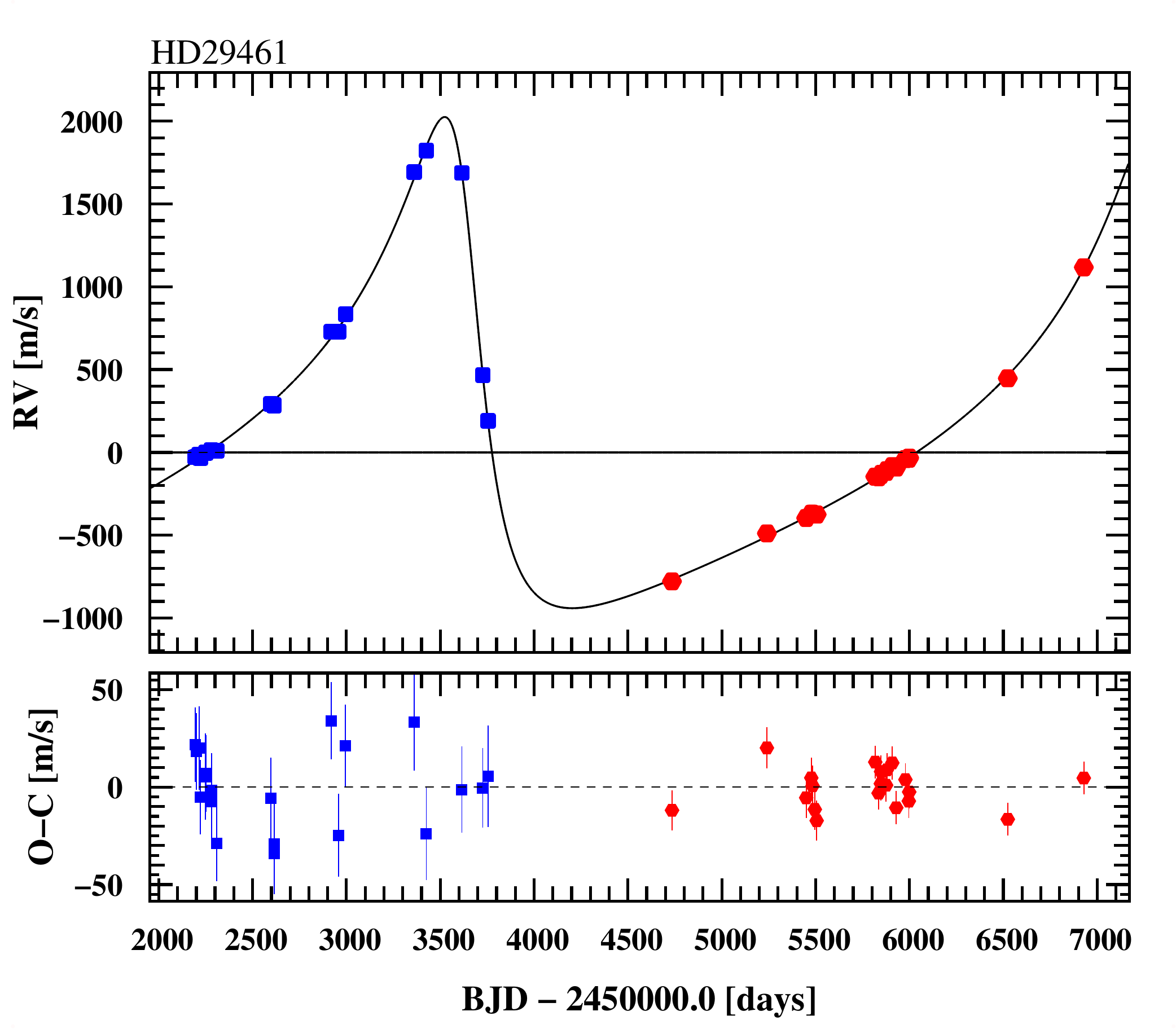}
\caption{Radial velocity curve of \object{HD29461}}
\label{fighd29461}
\end{figure}

\section{\object{HD175225}} 

\object{HD175225} is a V = 5.5 mag G9 subgiant located at 26.1 parsec from our Sun. 
\citet{2001AJ....121.3254T} made a direct detection of a companion at 1.1 arcsec (28.7 AU) 
using the adaptive optics system at Mount Wilson Observatory. They estimated from 
the V-I color that the companion corresponds to an M2 dwarf. Considering the separation, the orbital period is expected 
to be about 150 years. 
For this target we obtained 38 CORAVEL, 38 ELODIE, and 16 {\sophie} measurements in a time span 
of 31.7 years. Radial velocities, displayed in Fig.~\ref{fighd175225}, show a long-term trend with curvature. 
The Keplerian fit is no constraint, but indicates an orbital period longer than 45 years and a companion heavier than 0.2 $M_{\odot}$ 
, which is fully compatible with the direct imaging.

\begin{figure}[h]
\centering
\includegraphics[width=8cm]{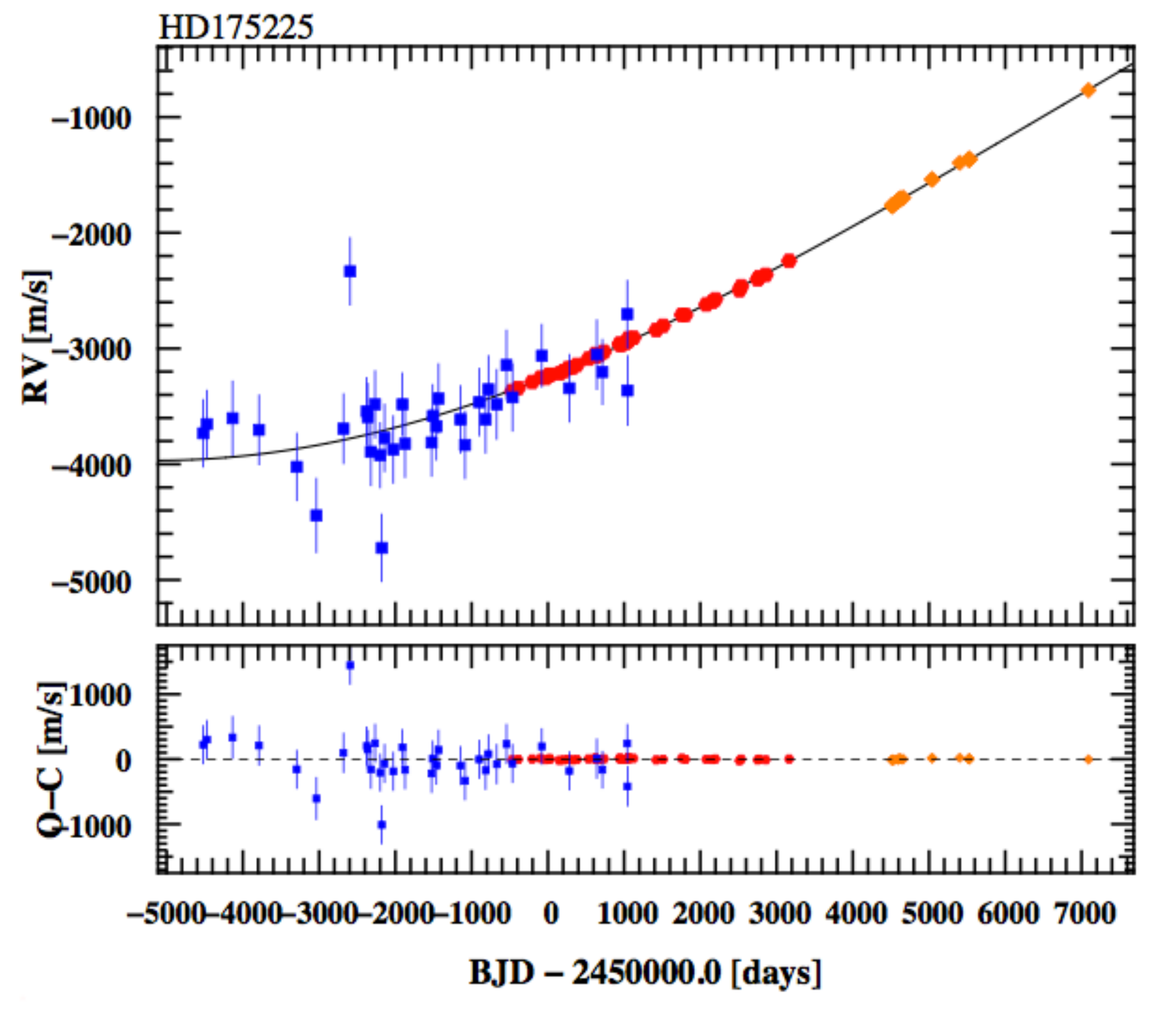}
\caption{Radial velocity curve of \object{HD175225} obtained with CORAVEL (blue), ELODIE (red), and SOPHIE (orange)}
\label{fighd175225}
\end{figure}

\end{document}